\documentclass[11pt,a4paper]{article}
\pdfoutput=1
\usepackage{jheppub}
\usepackage{bm}
\usepackage{mathrsfs}
\usepackage{slashed}

\newcommand{\downarrowtail}{\downarrow}

\usepackage{arydshln}

\usepackage[utf8]{inputenc}

\usepackage{graphicx}
\graphicspath{ {./figures/} }

\usepackage{comment}
\usepackage{subfigure}

\usepackage{amsmath} 

\allowdisplaybreaks

\newcommand\vek[1]{\bm{#1}}
\newcommand\he[1]{#1^\dagger}
\newcommand\adj[1]{\overline{#1}}
\newcommand\imag{\text{i}}
\newcommand\gr[1]{\mathrm{#1}}
\newcommand\La{\mathscr{L}}
\newcommand\dd{\mathop{\text d}\nolimits}
\newcommand\sumint[1]{\int\kern-1.5em\sum\nolimits_{#1}}
\newcommand\braced[1]{\{#1\}}
\newcommand\mufourpiT[1]{\left(\frac\Lambda{4\pi T}\right)^{#1\epsilon}}
\newcommand\OO{\mathcal{O}}

\newcommand{\MSbar}{$\overline{\text{MS}}$}
\newcommand\td{\text{3d}}
\newcommand\fd{\text{4d}}
\newcommand\RE{\text{Re}}
\newcommand\IM{\operatorname{Im}}
\newcommand\de{\partial}

\newcommand\eps{\epsilon}
\newcommand\mutwom[1]{\left(\frac\Lambda{2m}\right)^{#1\eps}}
\newcommand\xifac{\left(\frac{e^\gamma\Lambda^2}{4\pi}\right)^\eps}

\begin{document}

\title{{Three-dimensional effective theories for the two Higgs doublet model at high temperature}}

\preprint{HIP-2018-6/TH}

\author[a,b]{Tyler Gorda,}
\emailAdd{tyler.gorda@virginia.edu}
\affiliation[a]{Department of Physics and Helsinki Institute of Physics, P.O.~Box 64, FI-00014 University 
of Helsinki, Finland}
\affiliation[b]{Department of Physics, University of Virginia, 382 McCormick Road,
Charlottesville, Virginia 22904-4714, USA}

\author[c,d]{Andreas Helset,}
\emailAdd{ahelset@nbi.ku.dk}
\affiliation[c]{Niels Bohr International Academy and Discovery Center, Niels Bohr Institute,
University of Copenhagen, Blegdamsvej 17, DK-2100 Copenhagen, Denmark}
\affiliation[d]{Department of Physics, Faculty of Natural Sciences, Norwegian University of Science 
and Technology, Høgskoleringen 5, N-7491 Trondheim, Norway}

\author[a]{Lauri Niemi,}
\emailAdd{lauri.b.niemi@helsinki.fi}

\author[a,e]{Tuomas V.I.~Tenkanen}
\emailAdd{tenkanen@itp.unibe.ch}
\affiliation[e]{Albert Einstein Center for Fundamental Physics, Institute for Theoretical Physics, University of Bern, 
Sidlerstrasse 5, CH-3012 Bern, Switzerland}

\author[a]{\\ and David J.~Weir.}
\emailAdd{david.weir@helsinki.fi}

\abstract{Due to the infrared problem of high-temperature field
  theory, a robust study of the electroweak phase transition (EWPT)
  requires use of non-perturbative methods. We apply the method of
  high-temperature dimensional reduction to the two Higgs doublet
  model (2HDM) to obtain three-dimensional effective theories that can
  be used for non-perturbative simulations. A detailed derivation of
  the mapping between the full four-dimensional and the effective
  three-dimensional theories is presented. The results will be used in
  future lattice studies of the 2HDM. In the limit of large mass
  mixing between the doublets, existing lattice results can be
  recycled. The results of such a study are presented in a
  companion paper.}

\maketitle

\section{Introduction}

As the search for new particles at collider experiments continues, the
full structure of the scalar sector remains an active subject of
theoretical study. Beyond-the-Standard-Model (BSM) theories assuming a
richer Higgs sector are motivated by unanswered phenomenological
questions in the Standard Model (SM) and also by cosmological
observations suggesting, amongst other things, the existence of
an unknown dark matter particle. One particularly interesting BSM
theory is the two Higgs doublet model (2HDM), which augments the SM
with an additional scalar doublet and predicts new bosons that could
have observable signatures at present particle accelerators
\cite{Branco:2011iw,Dorsch:2014qja,Dorsch:2016tab}. In particular,
perturbative studies of the 2HDM at high temperatures suggest that it
may be possible to explain the observed matter/antimatter asymmetry by
means of electroweak baryogenesis (EWBG)
\cite{Basler:2016obg,Dorsch:2017nza}.

EWBG is a mechanism for generating an excess of baryonic matter during the electroweak phase transition (EWPT) via non-perturbative sphaleron processes near the bubble walls that form during a first-order phase transition \cite{Kuzmin:1985mm}. However, it is widely known from lattice simulations performed in the 1990's that the EWPT in the SM with a physical Higgs mass of 125 GeV is a crossover transition instead of first order, ruling out EWBG in the minimal SM \cite{Kajantie:1995kf,Kajantie:1996mn,Gurtler:1997hr,Csikor:1998eu}. It has also been demonstrated that another necessary ingredient for baryogenesis, CP violation, is too weak in the SM \cite{Gavela:1994dt,Brauner:2011vb,Brauner:2012gu}. However, models with multiple scalar doublets provide a mechanism for CP violation beyond that of the CKM matrix via a mixing term between the doublet fields \cite{Basler:2016obg,Basler:2017uxn}, making the 2HDM a viable candidate for the realization of EWBG. A thorough study of the nature of the EWPT in the 2HDM could thus provide insight on both the phenomenology of the model and the cosmological question of matter/antimatter asymmetry. 

First-order phase transitions at the electroweak scale are also a
source of gravitational waves, peaked at a characteristic frequency
given by the bubble radius, which would be in the mHz range today
\cite{Caprini:2015zlo,Weir:2017wfa}. This is within the sensitivity
region of LISA, so if the phase transition were strong enough,
its existence and properties could be probed through the
gravitational wave power spectrum it left behind
\cite{Hindmarsh:2017gnf}. Studies of gravitational waves from the 2HDM
have been carried out in the past in Refs.~\cite{Kakizaki:2015wua,Dorsch:2016nrg,Huang:2017rzf}. With the results of this paper, we will
  facilitate an improvement in the precision of these investigations.

Frequently, properties of the EWPT are studied in terms of a
perturbative finite-$T$ effective potential
\cite{Carrington:1991hz,Fromme:2006cm,Cline:2011mm,Dorsch:2013wja,Haarr:2016qzq,Alanne:2016wtx,Vaskonen:2016yiu,Marzola:2017jzl,Bernon:2017jgv}. This
approach contains a serious disadvantage: It is well-known that in
perturbation theory the symmetric phase is associated with disastrous
infrared (IR) problems \cite{Linde:1980ts,Gross:1980br}. Yet, in order
to find the critical temperature in perturbation theory---from the
condition that the symmetric and broken minima are
degenerate---information about the value of the potential at the
broken minimum, as well as the value of the potential at the origin,
is required. While the former quantity can be determined, as at
sufficiently large field values perturbation theory is applicable, the
latter quantity cannot be computed due to the non-perturbative nature
of the symmetric phase. This means that an accurate determination of
the critical temperature of the phase transition---as well as some
other thermodynamic quantities---is beyond the scope of perturbation
theory.  The determination of these quantities in perturbation theory
is always inaccurate without information about the behavior of the
potential near the origin. For this reason, reliable determination of
these quantities requires use of non-perturbative methods: in
practice, lattice Monte Carlo simulations. In the non-perturbative
studies of the SM in Ref.~\cite{Kajantie:1995kf}, considerable deviations
from perturbative computations of the effective potential were found
at small field values, and a more recent comparison in Ref.~\cite{Laine:2012jy}
reports an enhancement of $\sim 50\%$ in the latent heat.

In this paper, we take a different approach from earlier
perturbative studies, carrying out a procedure known as
finite-temperature dimensional reduction (DR), explained in detail in
Ref.~\cite{Kajantie:1995dw}, to construct effective three-dimensional
theories for the 2HDM. While the construction of the effective
theories is perturbative in nature, the DR procedure automatically
implements the required resummations for IR-sensitive contributions
\cite{Appelquist:1974tg,Braaten:1995cm,Kajantie:1995dw}. Each of these
theories can readily be studied non-perturbatively on the
lattice in an IR-safe manner. In particular, we describe a mapping to
a SM-like effective theory for which a non-perturbative
study is straightforward by making use of the lattice results of
Ref.~\cite{Kajantie:1995kf}. An application of this method is
presented in a companion paper \cite{Andersen:2017ika}, where we
present the results of parameter-space scans. Technical details
of the required calculations, as well as generalized parameter
mappings to the effective theories, are presented in the paper at
hand.

Despite the fact that DR and lattice methods in the context of the
EWPT have been successfully applied to the SM
\cite{Farakos:1994kx,Kajantie:1995dw} (for which the pressure has been computed using DR in Refs.~\cite{Gynther:2005dj,Gynther:2005av}) and the minimal supersymmetric
standard model (MSSM)
\cite{Laine:1996ms,Cline:1996cr,Laine:1998qk,Losada:1998at,Laine:2000rm}, these
methods are not widely used for BSM models with an extended scalar
sector. Our hope is that this detailed work will make the DR approach
more transparent, as it is a valuable alternative to the widely used,
fully perturbative method. Technical details presented in the
appendices of this work have also recently been used in
Ref.~\cite{Niemi:2018asa}, where DR has was applied to the SM
augmented with a real triplet, previously studied perturbatively in
Ref.~\cite{Patel:2012pi}. Similar techniques are currently being
applied to the real-singlet extension of the SM as well, while this model has already been analyzed---in
limited regions of parameter space---with the three-dimensional
approach in Ref.~\cite{Brauner:2016fla} (for perturbative analyses of the
EWPT in the singlet extension, see
Refs.~\cite{Beniwal:2017eik,Chen:2017qcz}). A compact and illuminating
review of the use of dimensional reduction has been presented in
Ref.~\cite{Laine:1997qm}. In the 2HDM, previous applications of DR can
be found in Refs.~\cite{Losada:1996ju,Andersen:1998br}: we extend
their results by fully including contributions from the $\gr{U(1)}$
gauge field and keeping certain parameters complex, which will allow
our parameter mapping to be applied to the CP-violating 2HDM.

This paper is organized as follows. In Section~\ref{sec:model} we
  introduce the model in Euclidean spacetime, while in
  Section~\ref{sec:dreft} we present the effective three-dimensional
  theories in a schematic form, before collecting together the main
  results of this paper: the matching relations between the full
  theory and the effective theories. The validity of our DR is also
  discussed. In Section~\ref{sec:discussion} we summarize key aspects
  of our study and outline future prospects. Some phenomenological
  implications of our study are discussed in the companion paper,
  Ref.~\cite{Andersen:2017ika}. In the interests of readability, many
  of the technical details of our derivation and results are relegated
  to the appendices.

\section{Description of the model}
\label{sec:model}

We will work in a $D=d+1=4-2\epsilon$ dimensional Euclidean spacetime.

\subsection{Full theory in four dimensions}
\label{sec:full4D}

The Lagrangian of the full theory reads 
\begin{equation}
\La=\La_\text{gauge}+\La_\text{ghost}+ \La_\text{gauge fixing}+\La_\text{fermion}+\La_\text{scalar}+\La_\text{Yukawa}+\delta\La + \La_\text{resummation},
\label{eq:classlag}
\end{equation}
where the gauge field, ghost, fermion, scalar and Yukawa sector Lagrangians are defined as follows: 
\begin{align}
\La_\text{gauge}={}&\frac14G^a_{\mu\nu}G^a_{\mu\nu}+\frac14F_{\mu\nu}F_{\mu\nu}+\frac14H^\alpha_{\mu\nu}H^\alpha_{\mu\nu}, \nonumber \\
\La_\text{ghost}={}&\partial_\mu\adj\eta^a D_\mu\eta^a+\partial_\mu\adj\xi\partial_\mu\xi+\partial_\mu\adj\zeta^\alpha D_\mu\zeta^\alpha, \nonumber \\
\La_\text{fermion}={}&\sum_A\left(\adj\ell_A\slashed D\ell_A+\adj e_A\slashed De_A+\adj q_A\slashed Dq_A+\adj u_A\slashed Du_A+\adj d_A\slashed Dd_A\right), \nonumber \\
\La_\text{scalar}={}&(D_\mu\phi_1)^\dagger (D_\mu\phi_1) +  (D_\mu\phi_2)^\dagger (D_\mu\phi_2) \nonumber \\ 
& \quad + \rho (D_\mu\phi_1)^\dagger (D_\mu\phi_2) + {\rho}^* (D_\mu\phi_2)^\dagger (D_\mu\phi_1) +  V(\phi_1, \phi_2), \nonumber \\
\La_\text{Yukawa} ={}& g_{Y} (\bar{q}_t \tilde\phi_2 t + \bar{t} \tilde\phi^\dagger_2 q_t).
\end{align}
UV counterterms in the modified minimal subtraction (\MSbar) scheme are contained in $\delta\La$: see Appendix~\ref{sec:4d_cts}.

In the gauge sector we have the $\gr{SU(2)}_L$, $\gr{U(1)}_Y$ and $\gr{SU(3)}_c$ gauge fields $A^a_\mu$, $B_\mu$, and $C^\alpha_\mu$ appearing inside the field strength tensors $G^a_{\mu\nu}$, $F_{\mu\nu}$ and $H^\alpha_{\mu\nu}$. The associated gauge couplings are $g$, $g'$, and $g_s$. The only ghost field of relevance for the phase transition is the $\gr{SU(2)}_L$ ghost $\eta^a$, as the $\gr{SU(3)}$ ghosts enter scalar correlation functions only at three-loop level and are heavily suppressed. Left-handed doublet and right-handed singlet lepton fields are denoted $\ell_A$ and $e_A$, with $A$ being the flavor index, while $q_A$ refers to left-handed doublet quark fields. $u_A$ and $d_A$ are right-handed singlet up- and down-type quark fields, respectively. The scalar sector consists of the doublet fields $\phi^i_1, \phi^i_2$ and the corresponding charge-conjugated fields $\tilde\phi_1 \equiv\imag\sigma_2\phi^*_1, \tilde\phi_2 \equiv\imag\sigma_2\phi^*_2$, where $\sigma_2$ is the second Pauli matrix. Finally, following the convention common in the literature, in the Yukawa sector we couple the top quark to $\phi_2$ only and neglect the other fermions\footnote{Models where individual fermions couple to both doublets are severely constrained due to resulting tree level flavor-changing neutral currents that have not been observed in experiments \cite{Branco:2011iw}.}. This is a good approximation in Type I 2HDMs---meaning that all fermions couple to one doublet only---where the other Yukawa couplings are small compared to other couplings in the theory. The relation $Q=I_3+\frac Y2$ between electric charge $Q$ and isospin
$I_3$ defines the hypercharge of the fields as follows: $Y_\ell=-1$,
$Y_e=-2$, $Y_q=\frac13$, $Y_u=\frac43$, $Y_d=-\frac23$, $Y_{\phi_1}=1$,
$Y_{\phi_2}=1$.

The scalar potential reads:
\begin{eqnarray}
\label{eq:scalar_potential}
V(\phi_1,\phi_2) &=& \mu^2_{11} \he\phi_1 \phi_1 + \mu^2_{22} \he\phi_2 \phi_2 + \mu^2_{12} \he\phi_1 \phi_2 + \mu^{2*}_{12} \he\phi_2 \phi_1 \nonumber \\
&+& \lambda_1 (\he\phi_1 \phi_1)^2 + \lambda_2 (\he\phi_2 \phi_2)^2 + \lambda_3 (\he\phi_1 \phi_1)(\he\phi_2 \phi_2)  + \lambda_4 (\he\phi_1 \phi_2)(\he\phi_2 \phi_1)  \nonumber \\
 &+& \frac{\lambda_5}{2} (\he\phi_1 \phi_2)^2  + \frac{\lambda^*_5}{2} (\he\phi_2 \phi_1)^2 + \lambda_6 (\he\phi_1 \phi_1)(\he\phi_1 \phi_2) + \lambda^*_6 (\he\phi_1 \phi_1)(\he\phi_2 \phi_1) \nonumber \\
&+& \lambda_7 (\he\phi_2 \phi_2)(\he\phi_2 \phi_1) + \lambda^*_7 (\he\phi_2 \phi_2)(\he\phi_1 \phi_2),
\end{eqnarray}
where the parameters $\mu^2_{11,22},\lambda_{1,2,3,4}$ are real and $\mu^2_{12}, \lambda_{5,6,7}$ are, in general, complex. Perturbative expansions of correlation functions, required for dimensional reduction, are organized in terms of the $\gr{SU(2)_L}$ gauge coupling $g$. We assume a power counting scheme where all mass parameters are heavy, i.e.,\  they scale as $\mu^2 \sim g^2 T^2$, and count powers of the quartic couplings as $\lambda \sim g^2$. In reality, however, the scalar couplings may be much larger in some regions of the parameter space. The gauge couplings $g,g',g_s$, as well as the top quark Yukawa coupling $g_Y$, are assumed to scale as $g$. The purpose of this schematic power counting is to facilitate the trunctuation of loop expansions, so that diagrams containing different types of fields are treated equally at each loop order.

The Lagrangian can be simplified by imposing a $Z_2$ symmetry. An exact $Z_2$ symmetry requires $\rho=\lambda_6=\lambda_7=\mu^2_{12}=0$, while a soft violation of the $Z_2$ symmetry is achieved with $\rho=\lambda_6=\lambda_7=0$, but $\mu^2_{12} \neq 0$ (see discussions in \cite{Ginzburg:2009dp,Ginzburg:2008kr}). In Ref.~\cite{Ginzburg:2005yw} it is described how a treatment of a true hard violation of the $Z_2$ symmetry is inconsistent without kinetic mixing terms with complex coupling $\rho$. In spite of this, we perform the dimensional reduction following Refs.~\cite{Losada:1996ju,Andersen:1998br} and set $\rho=0$ without imposing the full $Z_2$ symmetry; rather, we keep $\lambda_6$ and $\lambda_7$ in our calculation for technical reasons elaborated in Section \ref{sec:matching}. However, when turning to numerical analysis, we restrict ourselves to the case of soft violation of the $Z_2$ symmetry.

Consistent construction of the effective theory requires thermal resummation in order to remove problematic contributions originating from two-loop integrals with mixed Matsubara $n \neq 0$ and $n = 0$ modes \cite{Arnold:1992rz,Kajantie:1995dw}. We implement this in $\La_\text{resummation}$ by adding and subtracting one-loop thermal masses, denoted by $\bar{\Pi}$, as well as a thermal mixing mass term $\bar{\Pi}_{12}$, to zero modes of the scalar fields $\phi_1, \phi_2$. Schematically
\begin{align}
\Big(m^2 + \bar{\Pi} \Big) \he\phi(0,\vek p) \phi(0,\vek p) -\bar{\Pi} \he\phi(0,\vek p) \phi(0,\vek p)
 = \underline{m}^2 \he\phi(0,\vek p) \phi(0,\vek p) - \bar{\Pi} \he\phi(0,\vek p) \phi(0,\vek p), 
\end{align} 
where the Euclidean four-momentum is defined as $P=(\omega_n,\vek p)$ with $\omega_n=2n\pi T$, and we use the notation $\underline{m} \equiv \sqrt{\vphantom{\big\}} m^2 + \bar{\Pi}}$. Temporal components of the gauge fields are treated similarly; their thermal masses are just the corresponding Debye masses. The terms with minus signs, $-\bar{\Pi}$, are treated as counterterm-like interactions; hence we refer them to as thermal counterterms, despite being UV-finite. The thermal masses are listed explicitly in Appendix~\ref{sec:thermal_masses}. Terms with one-loop-resummed masses $\underline{\mu}^2$ and $+m^2_D, +{m_D'}^2$ contribute to propagators. This procedure is done both for the doublets $\phi_1, \phi_2$ and the gauge field temporal components $A^a_0,B_0$. The temporal gluon field $C^\alpha_0$ does not require resummation at order $O(g^4)$.

\subsection{Relations to physical quantities}
\label{sec:physical_parameter_rel}

We relate the Lagrangian parameters of the 2HDM to physical parameters at tree level, parameterizing the complex Higgs fields as
\begin{align}
    \phi_{1} &= \left( 
    \begin{array}{c}
    \phi_1^{+} \\
    \frac{1}{\sqrt{2}} \left( \rho_1 + i \eta_1 \right)
    \end{array}
    \right),\\
    \phi_{2} &= \left( 
    \begin{array}{c}
    \phi_2^{+} \\
    \frac{1}{\sqrt{2}} \left( \rho_2 + i \eta_2 \right)
    \end{array}
    \right).
\end{align}
In this section---and for the numerical analysis of the companion paper \cite{Andersen:2017ika}---we explicitly discard the $Z_2$ hard-breaking couplings $\lambda_6,\lambda_7$ from our scalar potential of Eq.~(\ref{eq:scalar_potential}).

We shall assume a CP-conserving vacuum\footnote{See Refs.~\cite{Ginzburg:2005yw,Barroso:2007rr,Tranberg:2012qu} for discussions on CP-breaking vacua.} and choose the vacuum expectation values (vevs) to be real, 
\begin{align}
\langle \phi_i \rangle = \frac{1}{\sqrt{2}}
\begin{pmatrix} 
0 \\ v_i 
\end{pmatrix}.
\end{align}
They satisfy the extremum conditions
\begin{equation}
    \frac{\partial V}{\partial \phi_i}\Big\vert_{\phi_i = \langle \phi_i \rangle}  = 0, \quad
    \frac{\partial V}{\partial \phi^\dagger_i}\Big\vert_{\phi_i = \langle \phi_i \rangle}  = 0,
\end{equation}
which lead to the following conditions for the mass parameters: 
\begin{align}
\mu _{11}^2 &= -\lambda _1 v_1^2 -\frac{1}{2} v_2^2 \lambda_{345}-\frac{v_2 \, \RE\mu _{12}^2}{v_1}, \\
\mu _{22}^2 &= -\lambda _2 v_2^2 -\frac{1}{2}v_1^2 \lambda_{345}-\frac{v_1 \, \RE\mu_{12}^2}{v_2}, \\
\IM\mu _{12}^2 &= -\frac{1}{2} v_1 v_2 \IM\lambda _5,
\end{align} 
where $\lambda_{345} \equiv \lambda_3+\lambda_4+\RE\,\lambda_5$. The vevs $v_1,v_2$ are constrained experimentally by the relation $v_1^2+v_2^2 = v^2 = (246\text{ GeV})^2$. The mixing of the two vevs is parameterized by the angle $\beta$, and we use the shorthand notation $t_\beta \equiv \tan\beta = v_2/v_1$. Furthermore, we shall also restrict our analysis to the region of parameter space where $\lambda_5$, and thus $\mu_{12}^2$, are real, and we choose to simplify the notation by denoting $\mu^2 \equiv -\RE\,\mu_{12}^2$.

Physical states are obtained from the $\phi_k^{\pm},\rho_k$ and $\eta_k$ by diagonalization and consist of two CP-even scalars $h, H_0$, a CP-odd pseudoscalar $A_0$ and the charged Higgses $H^{\pm}$. Three of the eight degrees of freedom are absorbed into Nambu-Goldstone bosons. The mass eigenstates are then related to $\phi_k^{\pm},\rho_k$, $\eta_k$ through
\begin{align}
h = -s_\alpha \rho_1 + c_\alpha \rho_2, \quad H_0 = - c_\alpha \rho_1 - s_\alpha \rho_2, \\
H^{\pm} = -s_\beta \phi_1^{\pm} + c_\beta \phi_2^{\pm}, \quad A_0 = -s_\beta \eta_1 + c_\beta \eta_2.
\end{align}
Here, $\alpha$ is defined as the mixing angle between the CP-even scalars, and we have introduced the shorthand notation $s_\alpha, c_\alpha, s_\beta, c_\beta \equiv \sin\alpha, \cos\alpha,\sin\beta,\cos\beta$. The quantity $c_{\beta-\alpha} \equiv \cos(\beta-\alpha)$ is phenomenologically important, as the choice $c_{\beta-\alpha} = 0$ corresponds to the alignment limit where $h$ couples to SM particles exactly like the Standard Model Higgs \cite{Gunion:2002zf}.

Physical masses are found by mass-matrix diagonalization and have been calculated in Refs. \cite{Branco:2011iw,Gunion:2002zf}, so we will not list them here. Inverting the eigenvalue relations allows us to write the Lagrangian parameters in terms of the masses $m_h,m_{H_0}, m_{A_0}, m_{H^{\pm}}$ and mixing parameters $t_\beta, c_{\beta-\alpha},\mu$, which are what we input into our parameter-space scans. These relations are listed in Appendix \ref{sec:parameter_constraints}. Precision tests of the 2HDM suggest that $m_{H^{\pm}}$ should be close to either $m_{H_0}$ or $m_{A_0}$ \cite{Gerard:2007kn,Grimus:2007if,Grimus:2008nb}. For the numerical analysis presented in the companion paper \cite{Andersen:2017ika}, we have chosen to set $m_{H^{\pm}} = m_{A_0}$, and have assumed that $h$ is the observed Higgs boson with mass $m_h = 125$ GeV.

For the gauge couplings and top Yukawa coupling, at tree-level
\begin{align}
g^2 &= g^2_0, \\
g'^2 &= \frac{g^2_0}{m^2_W}(m^2_Z - m^2_W), \\
g^2_{Y} &= \frac{g^2_0}{2} \frac{m^2_t}{m^2_W \sin^2\beta },
\end{align}
where we have denoted $g^2_0 \equiv 4\sqrt{2} G_f m^2_W$, with $G_f$ being the Fermi constant related to the lifetime of the muon. Similarly to the scalar couplings in Appendix~\ref{sec:params_tree}, we identify these as the renormalized parameters at \MSbar\ scale $\Lambda=m_Z$ ($\Lambda = m_t$ for the top Yukawa), neglecting the effects of vacuum renormalization on the \MSbar\ parameters. However, one-loop vacuum renormalization contributes at the same order in our power counting as two-loop dimensional reduction for doublet mass parameters and may have a sizeable effect on our results concerning the phase transition. We will address the numerical impact of zero-temperature renormalization in a future work. In the special case of the inert doublet model, one-loop vacuum renormalization calculations can be found in Ref.~\cite{Laine:2017hdk}.

Tree-level stability \cite{Klimenko:1984qx,Ginzburg:2005yw,Deshpande:1977rw,Kastening:1992by,Gunion:2002zf,Haber:1994mt} and unitarity \cite{Akeroyd:2000wc,Ginzburg:2003fe,Ginzburg:2005dt} requirements set additional constraints on the potential parameters. The relevant equations are listed in Appendix \ref{sec:parameter_constraints}. It has recently been pointed out that loop corrections to the stability conditions in the 2HDM are of importance \cite{Staub:2017ktc}; we plan to account for these in future work.

\section{Dimensional reduction and effective field theories}
\label{sec:dreft}

Physically, dimensional reduction is made possible by the fact that in
thermal equilibrium, the four-dimensional fields can be expressed in
terms of three-dimensional Matsubara modes that generate thermal
masses proportional to $\pi T$, a scale which we shall refer to as
``superheavy'' (see Table~\ref{tab:dr}). This causes all non-zero modes to decouple from long-distance physics at high temperatures. The remaining fields in the effective theory are just the bosonic zero modes.

\begin{table}[h]
\centering
\resizebox{\textwidth}{!}{%
\renewcommand{\arraystretch}{1.75} 

\begin{tabular}{cccccc}
  \multicolumn{6}{l}{\textsl{Start: \textbf{Two Higgs doublet model}}} \\
  \hline
\textbf{Name} & \textbf{Scale of validity} & \textbf{Dimensions} & \textbf{Lagrangian} & \textbf{Fields (excl. ghosts)} &
  \textbf{Parameters} \\
  \hline

  ``Superheavy'' & $\pi T$ & 4 & $\mathcal{L}_\text{full} \equiv
  \mathcal{L}$  (\ref{eq:classlag}) &
  $G_{\mu\nu},F_{\mu\nu},H_{\mu\nu},\phi_{1,2},\text{fermions}$ &
  $\mu_{11}^2,\mu_{12^2},\mu_{22}^2,g_{Y},g,g^{\prime} ,\lambda_{1}\cdots\lambda_{7}$ \\
  \multicolumn{6}{c}{$\Big\downarrow$ \textsl{Integrate out $n \neq 0$
      modes and fermions}} \\
  ``Heavy'' & $g T$ & 3 & $\mathcal{L}^{(3)}$ (\ref{eq:3d_heavy}) &
  $G_{rs},F_{rs},(H_{rs}),A_0,B_0,C_0,\phi_{1,2}$ &
  $m_D, m_D',m_D'',\mu^2_{11,3},\mu^2_{12,3},\mu^2_{22,3},$ \\
   & &  & & &
  $g_3,g_3^{\prime},\lambda_{1,3}\cdots\lambda_{7,3},h_1\cdots h_6$ \\
  \multicolumn{6}{c}{$\Big\downarrow$ \textsl{Integrate out temporal
      scalars}} \\
  ``Heavy'' & $gT$ & 3 & $\mathcal{\bar{L}}^{(3)}$
  (scalar part: \ref{eq:lag_3d_scalars}) &
  $G_{rs},F_{rs},(H_{rs}),\phi_{1,2}$ & $\bar{\mu}^2_{11,3},\bar{\mu}^2_{12,3},\bar{\mu}^2_{22,3}, \bar{g}_3, \bar{g}_3^{\prime},
  \bar{\lambda}_{1,3} \cdots \bar{\lambda}_{7,3}$ \\
  \hdashline
    \multicolumn{6}{l}{$\downarrowtail$ \textsl{\textbf{3D 2HDM} $\Rightarrow$
      new 3D simulations required}} \\
  \hdashline
  \multicolumn{6}{c}{$\Big\downarrow$ \textsl{Diagonalize theory at
      heavy scale}} \\
  ``Heavy'' & $gT$ & 3 & $\mathcal{\tilde{L}}^{(3)}$
  (scalar part: \ref{eq:3d_diagonal}) &
  $G_{rs},F_{rs},(H_{rs}),\theta,\phi$ & $\widetilde{\mu}^2_{\phi},\widetilde{\mu}^2_{\theta} \bar{g}_3, \bar{g}_3^{\prime},
  \widetilde{\lambda}_{1} \cdots \widetilde{\lambda}_{7}$ \\  
  \multicolumn{6}{c}{$\Big\downarrow$ \textsl{Where possible: integrate out heavy
        second doublet}} \\
``Light'' & $g^2 T$ & 3 & $\hat{\mathcal{L}}^{(3)}$ (\ref{eq:final_3d_lag})&
  $G_{rs},F_{rs},(H_{rs}),\phi$ & $\hat{\mu}_3^2,\hat{g}_3,\hat{g}_3',\hat{\lambda}_3$ \\
  \hline
    \multicolumn{6}{l}{\textsl{Finish: \textbf{3D minimal Standard Model}
      $\Rightarrow$ existing 3D simulations available}}
  \end{tabular}
}
\caption{\label{tab:dr} Dimensional reduction of the 2HDM into effective three-dimensional theories. Couplings of the effective theories are functions of the couplings of the full theory and are determined by the matching procedure. The last step is possible in the presence of a large $\mu^2_{12}$ mixing term.}
\end{table}

In practice, DR is performed by matching parameters of the three-dimensional theory to those of the full theory so that the long-distance Green's functions match. This requires perturbative calculations of correlation functions in the four-dimensional theory to a given accuracy and results in matching relations for parameters and fields in the effective theory. For DR, we apply renormalization in the \MSbar\ scheme; details can be found in Appendix~\ref{sec:4d_cts}.

The temporal components of the gauge fields are treated in the
three-dimensional theory as additional scalar fields with masses of
order $gT$: a mass scale we shall refer to as ``heavy''. These can be
integrated out as well to obtain a theory valid at the ``light'' scale
$g^2 T$, and information about the heavy and superheavy scales is then
contained in the fields and couplings of this effective theory. The
theory at the light scale is used to describe the equilibrium
thermodynamics of the full theory, as near the critical temperature thermal
corrections make at least one of the scalar mass parameters light. In
the presence of sizable mass mixing between the scalar doublets, we
may further integrate out one doublet for an even simpler effective theory; this
is described in Section~\ref{sec:SM-like}.

We emphasize that the three-dimensional theories are super-renormalizable and hence lattice simulations can reasonably be performed (see Ref.~\cite{Farakos:1995dn}), in contrast to performing direct simulations of the full four-dimensional theory. In addition, combining the perturbative method of DR with non-perturbative three-dimensional simulations is very efficient, as the DR procedure is free of IR problems and can be performed accurately using perturbation theory, while the latter is used to study the dynamics of the light bosonic modes, which are the source of IR problems in perturbative studies.

We work in Landau gauge, as this choice significantly simplifies many diagrammatic calculations. However, we highlight that to the order $O(g^4)$ that we work in, the parameters of the effective theory---in terms of $T$ and other physical quantities---are independent of the choice of gauge. This can be verified by performing the calculation in general covariant gauge and explicitly verifying the cancellation of the gauge parameter between terms from field normalization and correlation functions. Thus, DR combined with lattice simulations can be used to study the EWPT in a manifestly gauge-invariant manner. For discussions on gauge dependence in perturbative studies of the EWPT, see Refs.~\cite{Laine:1994zq,Laine:1994bf,Patel:2011th}.

\subsection{Effective theories at the heavy scale}

We denote the fields of the effective theories with the same symbols as those of the
four-dimensional theory, but emphasize that their normalization is different and will affect the mapping between the full and effective theories. For a generic field, the relation between the four-dimensional and three-dimensional fields reads \cite{Kajantie:1995dw}
\begin{equation}
\psi_\td^2 = \frac{1}{T}\big[1 + \Pi_{\psi}'(0) - \delta Z_\psi \big] \psi_\fd^2,
\label{fieldmatching}
\end{equation}
where $\Pi_{\psi}(P)$ is the self-energy of the field, a prime denotes
a derivative with respect to $P^2$, and $\delta Z_\psi$ is the field
renormalization counterterm.

The effective-theory gauge couplings are denoted by $g_3$ and $g'_3$. The Lagrangian of the first effective theory (again in Landau gauge) has the schematic form
\begin{equation}
\label{eq:3d_heavy}
\La^{(3)}=\La^{(3)}_\text{gauge}+\La^{(3)}_\text{ghost}+\La^{(3)}_\text{scalar}+\La^{(3)}_\text{temporal}+\delta \La^{(3)}.
\end{equation}
We include the $\gr{SU(2)}_L$ and $\gr{U(1)}_Y$ gauge fields in the gauge sector part,
\begin{equation}
\label{eq:gauge_3d}
\La^{(3)}_\text{gauge}=\frac14G^a_{rs}G^a_{rs}+\frac14F_{rs}F_{rs},
\end{equation}
where only spatial Lorentz indices are summed over. The spatial $\gr{SU(3)}_c$ gluon fields can be neglected at $O(g^4)$. 
 
The form of $\La^{(3)}_\text{scalar}$ is the same as in the four-dimensional theory, but we denote the couplings with an additional subscript, emphasizing that they are couplings of a three-dimensional theory. Furthermore, as a consequence of broken Lorentz symmetry in the temporal direction it is necessary to introduce additional scalar fields in the effective theory. These arise from the temporal components of gauge fields, hence we denote them by $A_0, B_0, C_0$ and call them temporal scalars. Their contribution reads
\begin{align}
\La^{(3)}_\text{temporal}={}&\frac12(D_rA^a_0)^2+\frac12m_D^2A^a_0A^a_0+\frac12(\de_rB_0)^2+\frac12m_D'^2B_0^2+\frac14\kappa_1(A^a_0A^a_0)^2+\frac14\kappa_2 B_0^4 \nonumber \\
&+\frac14\kappa_3 A^a_0A^a_0B_0^2+h_{1}\he\phi_1\phi_1 A^a_0A^a_0+h_{2}\he\phi_1\phi_1 B_0^2+h_{3} B_0\he\phi_1 A^a_0\sigma^a\phi_1 \nonumber \\
&+h_{4}\he\phi_2\phi_2 A^a_0A^a_0+h_{5}\he\phi_2\phi_2 B_0^2+h_{6}B_0\he\phi_2 A^a_0\sigma^a\phi_2 \nonumber \\
&+ \delta_1 \he\phi_1 \phi_2 A^a_0A^a_0 + \delta^*_{1} \he\phi_2 \phi_1 A^a_0A^a_0 + \delta_{2} \he\phi_1 \phi_2 B^2_0 + {\delta^*_{2}} \he\phi_2 \phi_1 B^2_0 \nonumber \\ 
&+ \delta_3 B_0\he\phi_1 A^a_0\sigma^a\phi_2 +  {\delta^*_3} B_0\he\phi_2 A^a_0\sigma^a\phi_1 \nonumber \\
& + \frac12(\partial_rC^\alpha_0)^2+\frac12m_D''^2C^\alpha_0C^\alpha_0+ \omega_3 C^\alpha_0C^\alpha_0 \he\phi_2\phi_2.
\end{align}
Here the (spatial) covariant derivative of an isospin triplet is $D_rA^a_0 = \de_r A^a_0 + g_3 \epsilon^{abc}A^b_rA^c_0$ and for the temporal gluon field 
we have used the usual derivative instead of the covariant derivative $D_rC^\alpha_0 = \partial_r C^\alpha_0 + g_s f^\alpha_{\phantom \alpha\beta\rho}C^\beta_rC^\rho_0$, as the operators $H^\alpha_{rs}H^\alpha_{rs}$, $(C^\alpha_0C^\alpha_0)^2$, $A^a_0A^a_0C^\alpha_0C^\alpha_0$ and $B_0^2C^\alpha_0C^\alpha_0$ have been discarded from the effective theory. Spatial gluons do not couple to the scalar fields, and self-interactions of temporal gluons and their interactions with other temporal scalars would have a very small contribution to quantities of interest, such as scalar mass parameters of the light scale effective theories.

The counterterm part $\delta \La^{(3)}$ plays an important role in determining relations between the continuum and lattice three-dimensional theories and is needed for the calculation of lattice counterterms \cite{Laine:1995np}. In a continuum three-dimensional theory with dimensional regularization, the one-loop correlation functions are finite, while two-loop contributions to self-energies contain UV divergences. The three-dimensional theory is super-renormalizable, and from the two-loop mass counterterms one can solve the exact running of the mass parameters in terms of the three-dimensional theory renormalization scale $\Lambda_3$ \cite{Farakos:1994kx}. The mass counterterms have been collected in Appendix~\ref{sec:mass_ct}.

Furthermore, since the scalar mass parameters can be close to zero near the phase transition, IR-sensitive contributions of the type $1/m^2$ need to be considered carefully. These appear in two-loop calculation of scalar two-point correlators. In order to perform the parameter matching, we apply a procedure analogous to the thermal resummation in the four-dimensional theory (see Section \ref{sec:full4D}) by adding and subtracting one-loop corrections from temporal scalar fields to fundamental scalar masses. Terms with plus signs contribute to the masses in scalar propagators, while terms with minus signs are treated as (counterterm-like) interactions, i.e.
\begin{align}
&\Big(\mu^2_{11,3} + \bar{\Pi}_{1,3} \Big) \he\phi_1 \phi_1 -\bar{\Pi}_{1,3} \he\phi_1 \phi_1 +\frac12 \Big(\mu^2_{22,3} + \bar{\Pi}_{2,3} \Big) \he\phi_2 \phi_2 - \frac12 \bar{\Pi}_{2,3} \he\phi_2 \phi_2 \nonumber \\
& = \underline{\mu}^2_{11,3} \he\phi_1 \phi_1 -\bar{\Pi}_{1,3} \he\phi_1 \phi_1 +\frac12 \underline{\mu}^2_{22,3} \he\phi_2 \phi_2 - \frac12 \bar{\Pi}_{2,3} \he\phi_2 \phi_2,
\end{align} 
where $\underline{m} \equiv \sqrt{m^2 + \bar{\Pi}}$. The effect of the new interactions is to cancel the IR-sensitive terms in the loop expansions, and the resulting matching relations are IR safe. Note that we do not need to include a counterterm interaction for the mixing mass parameter $\mu^2_{12,3}$, as the one-loop correction from the temporal scalar fields is of higher order. Explicit expressions for these mass corrections are given in Appendix~\ref{sec:thermal_masses}. 

The temporal scalar masses (Debye masses) are of the order $\sim g T$ and are thus safe to treat perturbatively. Following Ref.~\cite{Kajantie:1995dw}, we integrate these out in a separate step of dimensional reduction, obtaining a theory where the scalar sector has the form 
\begin{align}
\label{eq:lag_3d_scalars}
\bar{\La}^{(3)}_\text{scalar}&= (D_r\phi_1)^\dagger (D_r\phi_1) +  (D_r\phi_2)^\dagger (D_r\phi_2)  + \bar{V}(\phi_1, \phi_2),
\end{align}
and the parameters are denoted with a bar as $\bar{g}_3, {\bar{g}'}_3, \bar{\mu}^2_{11,3},$ etc. The gauge sector is as in Eq.~(\ref{eq:gauge_3d}). With the lattice-continuum relations presented in Refs.~\cite{Laine:1995np,Laine:1997dy}, this theory is readily studied non-perturbatively on the lattice using Monte Carlo simulations. 

\subsection{SM-like effective theory for the 2HDM}
\label{sec:SM-like}

In the limit of a large mass-mixing term $\mu^2_{12}$, we may simplify the effective theory of Eq.~(\ref{eq:lag_3d_scalars}) further by noticing that the phase transition takes place close to the point where the mass matrix has a zero eigenvalue, and in the diagonal basis the other mass parameter is then generically heavy. By performing a unitary transformation (see Appendix~\ref{sec:diagonalization}), one can remove the mixing mass term, and the resulting theory is given by
\begin{align}
\label{eq:3d_diagonal}
\widetilde{\La}^{(3)}_\text{scalar, diagonal}&= (D_r\phi)^\dagger (D_r\phi) +  (D_r\theta)^\dagger (D_r\theta)  + \widetilde{V}(\phi, \theta),
\end{align}
where the scalar potential reads
\begin{align}
\widetilde{V}&(\phi,\theta) = \widetilde{\mu}^2_{\phi} \he\phi \phi + \widetilde{\mu}^2_{\theta} \he\theta \theta  + \widetilde{\lambda}_1 (\he\phi \phi)^2 + \widetilde{\lambda}_2 (\he\theta \theta)^2 + \widetilde{\lambda}_3 (\he\phi \phi)(\he\theta \theta)  + \widetilde{\lambda}_4 (\he\phi \theta)(\he\theta \phi)  \nonumber \\
&+ \frac{\widetilde{\lambda}_5}{2} (\he\phi \theta)^2  + \frac{\widetilde{\lambda}^*_5}{2} (\he\theta \phi)^2 + \widetilde{\lambda}_6 (\he\phi \phi)(\he\phi \theta) + \widetilde{\lambda}^*_6 (\he\phi \phi)(\he\theta \phi)
+ \widetilde{\lambda}_7 (\he\theta \theta)(\he\theta \phi) + \widetilde{\lambda}^*_7 (\he\theta \theta)(\he\phi \theta),
\end{align}
and $\phi$ and $\theta$ are the light and heavy doublets, respectively. Note that in general the diagonalization procedure generates non-zero couplings $\widetilde{\lambda}_6$ and $\widetilde{\lambda}_7$ even in the case of a softly broken $Z_2$-symmetry.

The heavy doublet $\theta$ can be integrated out in a similar fashion as the temporal scalars. This leads to a final effective theory which has the same form as the effective theory constructed for the SM in Refs.~\cite{Farakos:1994kx,Kajantie:1995dw}:
\begin{align}
\label{eq:final_3d_lag}
\hat{\La}^{(3)}&= \frac14G^a_{rs}G^a_{rs}+\frac14F_{rs}F_{rs} + (D_r\phi)^\dagger (D_r\phi)  + V(\phi),
\end{align}
where the couplings are denoted with a hat as $\hat{g}_3, {\hat{g}'}_3$, and
\begin{align}
V(\phi) &= \hat{\mu}_3^2 \he\phi \phi  + \hat{\lambda}_3 (\he\phi \phi)^2. 
\end{align} 
This method of three-step DR is analogous to that of Ref.~\cite{Laine:1996ms} in the MSSM. Couplings are RG invariant, and the mass parameter runs at two-loop order. Due to super-renormalizability, the running of $\hat{\mu}_3^2$ can be solved exactly from two-loop mass renormalization, and the corresponding $\beta$ function receives no additional corrections at higher loop orders.

In certain regions of parameter space, it is possible for both doublets to be light in the vicinity of the electroweak phase transition, in which case the final three-dimensional effective theory is given by Eq.~(\ref{eq:lag_3d_scalars}). Non-perturbative studies in this theory require simulations with two dynamical doublets and are beyond the scope of our current study. Instead, we shall now focus on the regions of parameter space where the second doublet is heavy and can be integrated out. In this case, we use the DR matching relations that map the four-dimensional theory to the effective three-dimensional theory of Eq.~(\ref{eq:final_3d_lag}), and recycle the existing non-perturbative results of Ref.~\cite{Kajantie:1995kf}. Non-perturbative effects related to the $\gr{U(1)}$ gauge field were neglected in the aforementioned study; however, a non-perturbative analysis with $\gr{U(1)}$ field is presented in Ref.~\cite{Kajantie:1996qd} and shows no significant difference from the case where only the $\gr{SU(2)}$ field is considered. In our study, we include effects of the $\gr{U(1)}$ sector in our parameter matching, but use the simpler results of Ref.~\cite{Kajantie:1995kf} to analyze the phase structure of the 2HDM.

In the final effective theory, one of the four parameters $\hat{g}_3, {\hat{g}'}_3, \hat{\mu}_3^2, \hat{\lambda}_3$ can be used to measure all the dimensionful quantities as well as to fix the RG scale for the mass parameter (the couplings are RG invariant). We follow Ref.~\cite{Kajantie:1995kf} and choose $\hat{g}_3$. Then, the dynamics is determined by the three dimensionless ratios
\begin{align}
z \equiv \frac{\hat{g}'^2_3}{\hat g^2_3}, \quad \quad \quad \quad 
y \equiv \frac{\hat \mu_3^2(\hat g^2_3)}{\hat g^4_3}, \quad \quad \quad \quad
x \equiv \frac{\hat \lambda_3}{\hat g^2_3}.
\end{align}
Properties of the phase transition, however, depend essentially on only one parameter: As a justifiable approximation, non-perturbative effects of the $\gr{U(1)}$ gauge field can be neglected by setting $z=0$, and in practice, $y \approx 0$ on the critical line, close to its leading order value. This means that the character of the transition is described only by the magnitude of the parameter $x$. Results from Monte Carlo simulations \cite{Kajantie:1995kf} show that for a first-order transition, $0 \lesssim x \lesssim 0.11$. The transition gets weaker as $x$ increases, and above $x \approx 0.11$ only a smooth crossover remains.

With a DR mapping between the four-dimensional 2HDM and the SM-like three-dimensional effective theory, we can scan the physical parameter space, searching for $x<0.11$ and $y=0$ to find regions of first order transitions and the corresponding critical temperatures. Results of such parameter-space scans are presented in the companion paper \cite{Andersen:2017ika}. Note that if $x<0$ for some physical input parameters, the three-dimensional theory is not bounded from below and simulations are not possible. This indicates that our DR procedure has broken down, either because of neglected higher-order corrections to the matching relations, or neglected dimension six (hereafter 6-dim.) or higher-dimensional operators.

\subsection{Matching of the parameters}
\label{sec:matching}

The recipe for obtaining the matching relations has been presented in Refs.~\cite{Kajantie:1995dw,Brauner:2016fla}. In the first step of the DR, i.e., when the superheavy scale is integrated out, matching relations are calculated up to $O(g^4)$ in our power counting. This accuracy requires one-loop accuracy for couplings and two-loop for mass parameters (see Appendix~\ref{sec:effectivepot} for computational details). One-loop $\beta$ functions are required to make the matching relations independent of the renormalization scale at $O(g^4)$.  
In the second step of DR, when integrating out the heavy scale, it is convenient and numerically reasonable to perform calculations to the same loop order as in the first step of DR.

Although the main motivation for DR is to facilitate non-perturbative simulations, the DR procedure is perturbative, and the validity of perturbation theory at each step of the DR should therefore be estimated. Perturbative errors arise from two sources: Firstly, there are higher-order corrections to the parameters of the effective theories. Secondly, higher-dimensional operators have been neglected in the effective theories. We discuss these higher-order operators in Section~\ref{sec:DR_validity}. In the parameter-space scans of the companion paper \cite{Andersen:2017ika}, first-order phase transitions are mainly found in the large-mass regime where some of the couplings are large; hence, is is particularly important to estimate the validity of the DR procedure. For the same reason, we expect the one-loop-corrected relations to physical quantities to be of importance.

In the presence of the mixing term $\mu^2_{12} \phi^\dagger_1 \phi_2$, the correlation functions should be calculated only after a proper diagonalization of the scalar potential. Such a diagonalization is described in Appendix \ref{sec:diagonalization} and generally induces complex Yukawa couplings to the top quark for both doublets. However, under the scaling assumption $\mu^2_{12} \sim g^2 T^2$ we may evaluate the correlation functions in the off-diagonal basis where couplings remain simple by treating the mixing term as an interaction and neglecting contributions beyond $O(g^4)$. The matching relations below are derived in this fashion. Justifying the validity of this approach is straightforward by performing the DR properly in the diagonal basis where generally $\lambda_6, \lambda_7$ are non-vanishing, and comparing the resulting 3d parameters. Apart from the Yukawa contributions, we have verified numerically that the off-diagonal computation works very well for $|\mu^2_{12}| \lesssim (400 \, \text{GeV})^2$ and that the error is negligible. This is the main reason we keep the $Z_2$-violating couplings $\lambda_6, \lambda_7$ explicit in the matching relations.

In the DR procedure, by using thermal-mass-resummed propagators and corresponding thermal counterterms, we are explicitly able to show that at two-loop level, products of zero-mode and non-zero mode contributions in the correlation functions vanish. Due to this cancellation, one could neglect the effect of the zero modes at two-loop level as only the non-zero modes contribute to the final result. However, keeping the zero modes and explicitly verifying this cancellation serves as a valuable cross-check of our calculations, even though it technically complicates computations of the correlation functions.

We generalize the dimensional reduction presented in the companion paper \cite{Andersen:2017ika} to a general CP-violating 2HDM containing the complex $\lambda_6, \lambda_7$ terms. Furthermore, the relations presented below fully incorporate the contributions from the $\gr{U(1)}$ sector, which have been partly neglected in previous DR studies \cite{Losada:1996ju,Andersen:1998br}.

We use the following notation:
\begin{align}
N_d &=2, \nonumber \\
N_f &=3, \nonumber \\
L_b &\equiv2\ln\Big(\frac{\Lambda}{T}\Big)-2[\ln(4\pi)-\gamma], \nonumber \\ 
L_f &\equiv L_b+4\ln2, \nonumber \\
c &\equiv \frac{1}{2}\bigg(\ln\Big(\frac{8\pi}{9}\Big) + \frac{\zeta'(2)}{\zeta(2)} - 2 \gamma \bigg),
\end{align}
where $\gamma$ is the Euler-Mascheroni constant.

\subsubsection{Integration over the superheavy scale}
\label{sec:superheavy_matching}
Matching relations for the first step of DR, leading to the theory in Eq.~(\ref{eq:3d_heavy}), are listed in this section. When running of the $O(g^2)$ part is accounted for using the $\beta$ functions presented in Appendix \ref{sec:4d_cts}, the matching relations are manifestly independent of the renormalization scale $\Lambda$ to the order $O(g^4)$, except for the relations for the Debye masses, which we only calculate at one-loop level as they only enter the construction of the final effective theories through loop effects.
\begin{align}
m_D^2={}&g^2T^2\bigg(\frac{4+N_d}{6}+\frac{N_f}{3}\bigg),\\
m'^2_D={}&g'^2T^2\bigg(\frac{N_d}{6}+\frac{5N_f}{9}\bigg),\\
m''^2_D={}&g_s^2T^2\bigg(1+\frac{N_f}{6}\bigg),\\
g_3^2={}&g^2(\Lambda)T\Big(1 +\frac{g^2}{(4\pi)^2}\bigg[\frac{44-N_d}{6}L_b+\frac{2}{3}-\frac{4N_f}{3}L_f\bigg]\Big),\\
g'^2_3={}&g'^2(\Lambda)T\Big(1 +\frac{g'^2}{(4\pi)^2}\bigg[-\frac{N_d}{6}L_b-\frac{20N_f}{9}L_f\bigg]\Big),\\
\kappa_1={}&T\frac{g^4}{16 \pi^2} \frac{16+N_d-4N_f}{3},\\
\kappa_2={}&T\frac{g'^4}{16\pi^2} \bigg(\frac{N_d}{3}-\frac{380}{81} N_f\bigg),\\
\kappa_3={}&T\frac{g^2g'^2}{16\pi^2}\bigg(2N_d-\frac{8}{3}N_f\bigg),\\
\notag
h_{1}={}&\frac{g^2(\Lambda)T}{4}\bigg(1+\frac{1}{(4\pi)^2}\bigg\{\bigg[\frac{44-N_d}{6}L_b+\frac{53}{6}-\frac{N_d}{3}-\frac{4N_f}{3}(L_f-1)\bigg]g^2+\frac{g'^2}{2} \\
& + 12\lambda_1  + 2 (2\lambda_3+\lambda_4) \bigg\} \bigg), \\
h_{2}={}&\frac{g'^2(\Lambda)T}{4}\bigg(1 +\frac{1}{(4\pi)^2}\bigg\{\frac{3g^2}{2}+\bigg[\frac{1}{2}-\frac{N_d}{6}\Big(2+L_b \Big)  -\frac{20N_f}{9}(L_f-1)\bigg]g'^2 \notag \\
& + 12\lambda_1  + 2 (2\lambda_3+\lambda_4) \bigg\} \bigg), \\
h_{3}={}&\frac{g(\Lambda)g'(\Lambda)T}{2}\bigg\{1+\frac{1}{(4\pi)^2}\bigg[-\frac{5+N_d}{6} g^2+ \frac{3-N_d}{6}g'^2+L_b\bigg(\frac{44-N_d}{12}g^2 -\frac{N_d}{12}g'^2\bigg)\notag \\
&-N_f(L_f-1)\bigg(\frac{2}{3}g^2+\frac{10}{9}g'^2\bigg) + 4 \lambda_1 + 2 \lambda_4 \bigg]\bigg\},\\
h_{4}={}&\frac{g^2(\Lambda)T}{4}\bigg(1+\frac{1}{(4\pi)^2}\bigg\{\bigg[\frac{44-N_d}{6}L_b+\frac{53}{6}-\frac{N_d}{3}-\frac{4N_f}{3}(L_f-1)\bigg]g^2+\frac{g'^2}{2} -6 g_{Y}^2\notag\\
& + 12\lambda_2  + 2 (2\lambda_3+\lambda_4) \bigg\} \bigg), \\
h_{5}={}&\frac{g'^2(\Lambda)T}{4}\bigg(1 +\frac{1}{(4\pi)^2}\bigg\{\frac{3g^2}{2}+\bigg[\frac{1}{2}-\frac{N_d}{6}\Big(2+L_b \Big)  -\frac{20N_f}{9}(L_f-1)\bigg]g'^2 - \frac{34}{3} g_{Y}^2 \notag \\
&  + 12\lambda_2  + 2 (2\lambda_3+\lambda_4) \bigg\} \bigg), \\
h_{6}={}&\frac{g(\Lambda)g'(\Lambda)T}{2}\bigg\{1+\frac{1}{(4\pi)^2}\bigg[-\frac{5+N_d}{6} g^2+ \frac{3-N_d}{6}g'^2+L_b\bigg(\frac{44-N_d}{12}g^2 -\frac{N_d}{12}g'^2\bigg)\notag \\
&-N_f(L_f-1)\bigg(\frac{2}{3}g^2+\frac{10}{9}g'^2\bigg) + 2 g_{Y}^2 + 4 \lambda_2 + 2 \lambda_4 \bigg]\bigg\},\\
\delta_1={}&  \frac{3}{2} \frac{g^2 T}{16\pi^2}(\lambda_6 + \lambda^*_7),\\
\delta_2={}& \frac{3}{2}\frac{{g'}^2 T}{16\pi^2}(\lambda_6 + \lambda^*_7),\\
\delta_3={}& \frac{g{g'} T}{16\pi^2}(\lambda_6 + \lambda^*_7),\\
\omega_3={}&-T\frac{1}{16 \pi^2} 2 g^2_s g^2_{Y},\\
\lambda_{1,3}={}&T\Big(\lambda_1(\Lambda) + \frac{1}{(4\pi)^2}\bigg[\frac{1}{8}\Big(3g^4 + {g'}^4 +2 g^2{g'}^2 \Big) -L_b \bigg(\frac{3}{16}\Big(3g^4 + {g'}^4 + 2 g^2{g'}^2 \Big) \nonumber  \\
& + \lambda^2_3 + \lambda_3 \lambda_4 + \frac{1}{2}\lambda^2_4 + \frac{1}{2}|\lambda_5|^2 +  6 |\lambda_6|^2 - \frac{3}{2}\Big(3g^2+{g'}^2 -8 \lambda_1 \Big) \lambda_1 \bigg) \bigg]\Big), \\
\lambda_{2,3}={}&T\Big(\lambda_2(\Lambda) + \frac{1}{(4\pi)^2}\bigg[\frac{1}{8}\Big(3g^4 + {g'}^4 +2 g^2{g'}^2 \Big)+ 3 L_f \Big(g^4_{Y} - 2\lambda_2 g^2_{Y} \Big) \nonumber  \\
& -L_b \bigg(\frac{3}{16}\Big(3g^4 + {g'}^4 + 2 g^2{g'}^2 \Big) + \lambda^2_3 + \lambda_3 \lambda_4 + \frac{1}{2}\lambda^2_4 + \frac{1}{2}|\lambda_5|^2 +  6 |\lambda_7|^2 \nonumber \\ 
& -\frac{3}{2}\Big(3g^2+{g'}^2 -8 \lambda_2 \Big) \lambda_2 \bigg) \bigg]\Big), \\
\lambda_{3,3}={}&T\Big(\lambda_3(\Lambda) + \frac{1}{(4\pi)^2}\bigg[\frac{1}{4}\Big(3g^4 + {g'}^4 -2 g^2{g'}^2 \Big) - 3 L_f \lambda_3  g^2_{Y} \nonumber  \\
& -L_b \bigg(\frac{3}{8}\Big(3g^4 + {g'}^4 - 2 g^2{g'}^2 \Big) + 2(\lambda_1 + \lambda_2)(3\lambda_3 + \lambda_4) + 2\lambda^2_3 +  \lambda^2_4 + |\lambda_5|^2 \notag \\
&+ 2( |\lambda_6|^2+ |\lambda_7|^2) + 8 \RE(\lambda_6 \lambda_7) - \frac{3}{2}\Big(3g^2+{g'}^2\Big) \lambda_3 \bigg) \bigg]\Big), \\
\lambda_{4,3}={}&T\Big(\lambda_4(\Lambda) + \frac{1}{(4\pi)^2}\bigg[ g^2{g'}^2  - 3 L_f \lambda_4  g^2_{Y}  \nonumber  \\
& -L_b \bigg(\frac{3}{2} g^2{g'}^2 + 2(\lambda_1 + \lambda_2)\lambda_4 + 2\lambda^2_4 + 4\lambda_3\lambda_4 + 4|\lambda_5|^2  \notag \\ 
& + 5( |\lambda_6|^2+ |\lambda_7|^2) + 2 \RE(\lambda_6 \lambda_7) - \frac{3}{2}\Big(3g^2+{g'}^2\Big) \lambda_4 \bigg) \bigg]\Big), \\
\lambda_{5,3}={}&T\Big(\lambda_5(\Lambda) + \frac{1}{(4\pi)^2}\bigg[  -3 L_f \lambda_5 g^2_{Y} 
-L_b \bigg( 2(\lambda_1 + \lambda_2 + 2 \lambda_3 + 3 \lambda_4)\lambda_5 \nonumber \\ 
&+ 5(\lambda_6 \lambda_6 + \lambda^*_7 \lambda^*_7) 
+ 2 \lambda_6 \lambda^*_7 - \frac{3}{2}\Big(3g^2+{g'}^2\Big) \lambda_5 \bigg) \bigg]\Big), \\
\lambda_{6,3}={}&T\Big(\lambda_6(\Lambda) + \frac{1}{(4\pi)^2}\bigg[  -\frac{3}{2} L_f \lambda_6 g^2_{Y} -L_b \bigg( 12 \lambda_1 \lambda_6 + (3\lambda_3 + 2\lambda_4) \lambda^*_7 \nonumber  \\
&+ \lambda_5\lambda_7 + (3\lambda_3 + 4\lambda_4)\lambda_6 + 5 \lambda_5 \lambda^*_6  - \frac{3}{2}\Big(3g^2+{g'}^2\Big) \lambda_6 \bigg) \bigg]\Big), \\
\lambda_{7,3}={}&T\Big(\lambda_7(\Lambda) + \frac{1}{(4\pi)^2}\bigg[  -\frac{9}{2} L_f  \lambda_7 g^2_{Y} -L_b \bigg( 12 \lambda_2 \lambda_7 + (3\lambda_3 + 2\lambda_4) \lambda^*_6 \nonumber  \\
&+ \lambda^*_5\lambda_6 + (3\lambda_3 + 4\lambda_4)\lambda_7 + 5 \lambda^*_5 \lambda^*_7  - \frac{3}{2}\Big(3g^2+{g'}^2\Big) \lambda_7 \bigg) \bigg]\Big).
\end{align}
The SM result for the three-dimensional mass parameter reads
\begin{align}
\Big(\mu^2_{22,3}\Big)_\text{SM} =& \mu^2_{22}(\Lambda) +\frac{T^2}{16}\Big(3g^2(\Lambda) + {g'}^2(\Lambda) + 4 g^2_Y(\Lambda) + 8 \lambda_2(\Lambda) \Big) \nonumber \\
& + \frac{1}{16\pi^2} \bigg\{ \mu^2_{22}\bigg( \Big(\frac{3}{4}(3g^2 + {g'}^2) - 6 \lambda_2 \Big)L_b - 3 g^2_Y L_f \bigg) \nonumber \\
& + T^2 \bigg( \frac{167}{96}g^4 + \frac{1}{288}{g'}^4 - \frac{3}{16}g^2{g'}^2 + \frac{1}{4}\lambda_2(3g^2+{g'}^2) \nonumber \\
& + L_b \Big( \frac{17}{16}g^4 - \frac{5}{48}{g'}^4 - \frac{3}{16}g^2{g'}^2 + \frac{3}{4}\lambda_2(3g^2+{g'}^2) - 6 \lambda^2_2 \Big) \nonumber \\
& + \frac{1}{T^2}\Big( c + \ln(\frac{3T}{\Lambda_{3d}}) \Big)\Big( \frac{39}{16}g^4_3 + 12g^2_3 h_{4} - 6 h^2_{4} + 9 g^2_3 \lambda_{2,3} - 12\lambda^2_{2,3} \nonumber \\
& -\frac{5}{16}{g'}^4_3 - \frac{9}{8}g^2_3 {g'}^2_3 - 2h^2_{5} - 3h^2_6 + 3 {g'}^2_3 \lambda_{2,3} \Big) \nonumber \\
& - g^2_Y \Big(\frac{3}{16}g^2 + \frac{11}{48}{g'}^2 + 2 g^2_s \Big) + (\frac{1}{12}g^4 + \frac{5}{108}{g'}^4)N_f \nonumber \\
& + L_f \Big( g^2_Y \Big(\frac{9}{16}g^2 + \frac{17}{48}{g'}^2 + 2 g^2_s - 3 \lambda_2 \Big) +\frac{3}{8}g^4_Y - (\frac{1}{4}g^4 + \frac{5}{36}{g'}^4 ) N_f \Big) \nonumber \\
& + \ln(2) \Big( g^2_Y \Big(-\frac{21}{8}g^2 - \frac{47}{72}{g'}^2 + \frac{8}{3} g^2_s + 9 \lambda_2 \Big) -\frac{3}{2}g^4_Y + (\frac{3}{2}g^4 + \frac{5}{6}{g'}^4 ) N_f \Big) \bigg) \bigg\}.
\end{align}
This result can also be found from Ref.~\cite{Kajantie:1995dw}, apart from the two-loop contributions involving ${g'}$, as it was assumed to scale as ${g'}\sim g^{3/2}$. In the 2HDM, full results for the scalar mass parameters read:
\begin{align}
\Big(\mu^2_{22,3}\Big)_\text{2HDM} =& \Big(\mu^2_{22,3}\Big)_\text{SM} + \frac{T^2}{12}\Big(2\lambda_3(\Lambda)+\lambda_4(\Lambda) \Big) \nonumber \\
& + \frac{1}{16\pi^2}\bigg\{  \mu^2_{11}\Big( -L_b(2\lambda_3 + \lambda_4) \Big) \nonumber \\
& + T^2 \bigg( \frac{5}{48} g^4 + \frac{5}{144} {g'}^4 + \frac{1}{24}(3g^2 + {g'}^2)( 2\lambda_3 + \lambda_4) \nonumber \\
&+ \frac{1}{T^2}\Big(c + \ln\big(\frac{3T}{\Lambda_{3}} \big) \Big)\Big( -\frac{1}{8}(3 g^4_3 + {g'}^4_3) + \frac{1}{2}(3g^2_3 + {g'}^2_3)(2\lambda_{3,3} + \lambda_{4,3}) \nonumber \\ 
& - 2 (\lambda^2_{3,3} + \lambda_{3,3} \lambda_{4,3} + \lambda^2_{4,3})  - 3 |\lambda_{5,3}|^2 \Big) \nonumber \\
&+ L_b \Big( -\frac{7}{32}g^4 - \frac{7}{96}{g'}^4
-\frac{1}{2}(\lambda_1+\lambda_2)(2\lambda_3+\lambda_4) \nonumber \\
& -\frac{5}{6}\lambda^2_3 - \frac{7}{12}\lambda^2_4 -\frac{5}{6} \lambda_3 \lambda_4 - \frac{3}{4}|\lambda_5|^2 +\frac{1}{8}(3g^2+{g'}^2) \big( 2\lambda_3 + \lambda_4 \big)  \Big) \nonumber \\
&+ \Big( - \frac{1}{4}g^2_Y \big(2\lambda_3 + \lambda_4 \big)  \Big)L_f  \bigg)  \nonumber \\
&-6 L_b \RE(\mu^2_{12}\lambda_7) + T^2 \bigg( L_b \Big( -\frac{15}{4}|\lambda_7|^2 -\frac{3}{4}|\lambda_6|^2 - \frac{3}{2}\RE(\lambda_6 \lambda_7) \Big) \nonumber \\
&+ \frac{1}{T^2}\Big(c + \ln\big(\frac{3T}{\Lambda_{3}} \big) \Big)\Big(-3|\lambda_{6,3}|^2  -9|\lambda_{7,3}|^2  \Big) \bigg) \bigg\}
\end{align}
and
\begin{align}
\mu^2_{11,3} =& \mu^2_{11}(\Lambda) +\frac{T^2}{16}\Big(3g^2(\Lambda) + {g'}^2(\Lambda) + 8 \lambda_1(\Lambda) +\frac{4}{3}\Big(2\lambda_3(\Lambda) + \lambda_4(\Lambda) \Big)  \Big) \nonumber \\
& + \frac{1}{16\pi^2} \bigg\{L_b \bigg( \Big(\frac{3}{4}(3g^2 + {g'}^2) - 6 \lambda_1 \Big)\mu^2_{11} - (2\lambda_3 + \lambda_4)\mu^2_{22} - 6 \RE(\lambda_6 \mu^{2*}_{12}) \bigg) \nonumber \\
& + T^2 \bigg( \frac{59}{32}g^4 + \frac{11}{288}{g'}^4 - \frac{3}{16}g^2{g'}^2 + \frac{1}{4}\lambda_1(3g^2+{g'}^2) +\frac{1}{24}(3g^2 + {g'}^2)( 2\lambda_3 + \lambda_4) \nonumber \\
& + L_b \Big( \frac{27}{32}g^4 - \frac{17}{96}{g'}^4 - \frac{3}{16}g^2{g'}^2 + \frac{1}{8}(3g^2+{g'}^2)(6\lambda_1 + 2\lambda_3 + \lambda_4) \nonumber \\ 
&-\frac{1}{2}(\lambda_1+\lambda_2)(2\lambda_3 + \lambda_4) - 6 \lambda^2_1 - \frac{5}{6}\lambda^2_3 - \frac{5}{6}\lambda_3 \lambda_4 - \frac{7}{12}\lambda^2_4 - \frac{3}{4}|\lambda_5|^2 - \frac{3}{4} |\lambda_7|^2 \nonumber \\ 
&- \frac{15}{4} |\lambda_6|^2 - \frac{3}{2}\RE(\lambda_6 \lambda_7) \Big) \nonumber \\
& + \frac{1}{T^2}\Big( c + \ln(\frac{3T}{\Lambda_{3}}) \Big)\Big( \frac{33}{16}g^4_3 + 12g^2_3 h_{1} - 6 h^2_{1} + 9 g^2_3 \lambda_{1,3} - 12\lambda^2_{1,3} \nonumber \\
& -\frac{7}{16}{g'}^4_3 - \frac{9}{8}g^2_3 {g'}^2_3 - 2h^2_{2} - 3h^2_{3} + 3 {g'}^2_3 \lambda_{1,3} + \frac{1}{2}(3g^2_3 + {g'}^2_3)(2\lambda_{3,3} + \lambda_{4,3}) \nonumber \\
&- 2 (\lambda^2_{3,3} + \lambda_{3,3} \lambda_{4,3} + \lambda^2_{4,3}) - 3 |\lambda_{5,3}|^2 - 3 |\lambda_{7,3}|^2 - 9 |\lambda_{6,3}|^2 \Big) \nonumber \\
& + (\frac{1}{12}g^4 + \frac{5}{108}{g'}^4)N_f + L_f \Big( -\frac{1}{4}g^2_Y(2\lambda_3 + \lambda_4)  - (\frac{1}{4}g^4 + \frac{5}{36}{g'}^4 ) N_f \Big) \nonumber \\
& + \ln(2) \Big( \frac{3}{2} g^2_Y \Big(2\lambda_3 + \lambda_4 \Big) + (\frac{3}{2}g^4 + \frac{5}{6}{g'}^4 ) N_f \Big) \bigg) \bigg\},
\end{align}
and finally 
\begin{align}
\mu^2_{12,3} =& \mu^2_{12}(\Lambda) + \frac{T^2}{4} \Big(\lambda^*_7(\Lambda) + \lambda_6(\Lambda) \Big) + \frac{1}{16\pi^2}\bigg\{ L_b \bigg(\Big(\frac{3}{4}(3g^2+{g'}^2) -\lambda_3 -2 \lambda_4 \Big) \mu^2_{12} \nonumber \\
& -3 \Big( \lambda_5 \mu^{2*}_{12} + \lambda_6 \mu^2_{11} + \lambda^*_7 \mu^2_{22}\Big)  \bigg) - \frac{3}{2}g^2_Y \mu^2_{12} L_f + T^2 \bigg( -(\frac{3}{8}\lambda_6+\frac{9}{8}\lambda^*_7) g^2_Y L_f \nonumber \\
& + \frac{1}{8}(3g^2+{g'}^2)(\lambda_6 + \lambda^*_7)
+ \frac{1}{8} L_b \Big[ 3(\lambda_6+\lambda^*_7)\Big(3g^2 + {g'}^2 -4(\lambda_3+\lambda_4)  \Big) 
 \nonumber \\  
&-12\lambda_5\lambda^*_6 - 12 \lambda_5 \lambda_7 - 24 \lambda_2 \lambda^*_7 - 24\lambda_1\lambda_6 \Big] \nonumber \\
&  + \frac{1}{T^2}\Big( c + \ln\big(\frac{3T}{\Lambda_{3}} \big) \Big) \Big( \frac{3}{2} (3 g_3^2+{g_3'}^2 ) (\lambda _{6,3}+\lambda _{7,3}^*)-3\lambda _{6,3}(2 \lambda_{1,3} + \lambda _{3,3} +\lambda _{4,3})\\ \nonumber
&-3\lambda _{7,3}^* (2 \lambda _{2,3} + \lambda _{3,3} + \lambda _{4,3} ) -3 \lambda _{5,3}(\lambda_{6,3}^*+ \lambda_{7,3}) \Big) \bigg) + \frac{9}{2}g_Y^2 \lambda^*_7 \ln(2) \bigg\}.
\end{align}
The renormalization scale in the three-dimensional theory, $\Lambda_{3}$, as well as other three-dimensional parameters, appear above as exact solutions of the RG equations for the mass parameters of the effective theory, and we emphasize that this running is separate from that of the full four-dimensional theory.

\subsubsection{Integrating out the temporal scalars}
\label{sec:heavy_matching}

Results for parameter matching when integrating out the temporal scalar fields are listed below.

\begin{align}
\bar{g}^2_3 =& g^2_3 \Big( 1 - \frac{g^2_3}{24\pi m_D} \Big), \\
\bar{g}'^2_3 =& g'^2_3, \\
\bar{\mu}^2_{11,3} =& \mu^2_{11,3} - \frac{1}{4\pi}\Big(3 h_{1} m_D +  h_{2} m_D' \Big) \nonumber \\
& + \frac{1}{16\pi^2} \bigg( 3g^2_3 h_{1} - 3 h^2_{1} - {h}^2_{2} - \frac{3}{2}{h}^2_{3} \nonumber \\
& + \Big(-\frac{3}{4}g^4_3 + 12 g^2_3 h_{1} \Big) \ln\Big(\frac{\Lambda_{3}}{2m_D} \Big) 
 - 6 h^2_{1}  \ln\Big(\frac{\Lambda_{3}}{2m_D + \mu_{11,3}} \Big) \nonumber \\
&  - 2 {h}^2_{2} \ln\Big(\frac{\Lambda_{3}}{2m_D' + \mu_{11,3}}
 \Big) - 3 {h}^2_{3} \ln\Big(\frac{\Lambda_{3}}{m_D+m_D'+ \mu_{11,3}}
 \Big) \nonumber \\
& + 2 \mu_{11,3} \Big( 3\frac{h^2_{1}}{m_D} + \frac{{h}^2_{2}}{m_D'} \Big)  + 2 \mu_{22,3} \Big( 3\frac{ h_{1}h_{4}}{m_D} + \frac{  {h}_{2}{h}_{5}}{m_D'} \Big) \bigg)
, \\ 
\bar{\mu}^2_{22,3} =& \mu^2_{22,3} - \frac{1}{4\pi}\Big(3 h_{4} m_D +  h_{5} m_D' + 8 \omega_3 m_D'' \Big) \nonumber \\
& + \frac{1}{16\pi^2} \bigg( 3g^2_3 h_{4} - 3 h^2_{4} - {h}^2_{5} - \frac{3}{2}{h}^2_{6} \nonumber \\
& + \Big(-\frac{3}{4}g^4_3 + 12 g^2_3 h_{4} \Big) \ln\Big(\frac{\Lambda_{3}}{2m_D} \Big) 
 - 6 h^2_{4}  \ln\Big(\frac{\Lambda_{3}}{2m_D + \mu_{22,3}} \Big) \nonumber \\
& - 2 {h}^2_{5} \ln\Big(\frac{\Lambda_{3}}{2m_D' +
   \mu^2_{22,3}} \Big) - 3 {h}^2_{6}
 \ln\Big(\frac{\Lambda_{3}}{m_D+m_D'+ \mu_{22,3}} \Big) \nonumber \\
& + 2 \mu^2_{22,3} \Big( 3\frac{h^2_{4}}{m_D} + \frac{{h}^2_{5}}{m_D'} \Big)  + 2 \mu_{11,3} \Big( 3\frac{ h_{4}h_{1}}{m_D} + \frac{ {h}_{5}{h}_{2}}{m_D'} \Big)  \bigg)
, \\
\bar{\mu}^2_{12,3} =& \mu^2_{12,3}, \\ 
\bar{\lambda}_{1,3} =& \lambda_{1,3} - \frac{1}{8\pi}\Big( \frac{3 h^2_{1}}{m_D} + \frac{ h_{2}^2 }{m_D'} + \frac{ h_{3}^2}{m_D+m_D'} \Big), \\
\bar{\lambda}_{2,3} =& \lambda_{2,3} - \frac{1}{8\pi}\Big( \frac{3 h^2_{4}}{m_D} + \frac{ h_{5}^2 }{m_D'} + \frac{ h_{6}^2}{m_D+m_D'} \Big), \\
\bar{\lambda}_{3,3} =& \lambda_{3,3} - \frac{1}{4\pi}\Big( \frac{3 h_{1}h_{4}}{m_D} + \frac{ h_{2} h_{5} }{m_D'} + \frac{ h_{3} h_{6}}{m_D+m_D'} \Big), \\
\bar{\lambda}_{4,3} =& \lambda_{4,3}, \\
\bar{\lambda}_{5,3} =& \lambda_{5,3}, \\
\bar{\lambda}_{6,3} =& \lambda_{6,3}, \\
\bar{\lambda}_{7,3} =& \lambda_{7,3}
\end{align}
As before, the logarithms correspond to the running of masses in the resulting theory. For the numerical analysis presented in the companion paper~\cite{Andersen:2017ika}, we have chosen the RG scale in the resulting 3d theory as $\Lambda_{3}' = \bar g^2_3$, and have verified that numerical uncertainties from scale variations in this step of DR are negligible.

\subsubsection{Integrating out the heavy second doublet}

Finally, we present matching relations for the SM-like effective theory described in Section \ref{sec:SM-like}. In order to integrate out the heavy doublet, we diagonalize the scalar Lagrangian by means of a unitary transformation. 
Relations between couplings of the off-diagonal and diagonalized theories are given in Appendix~\ref{sec:diagonalization}.

\begin{align}
\hat{g}^2_3 =& {\bar{g}}^2_3 \Big( 1 - \frac{{\bar{g}}^2_3}{48\pi \widetilde{\mu}_\theta} \Big), \\
\hat{g}'^2_3 =& \bar{g}'^2_3 \Big( 1 - \frac{\bar{g}'^2_3}{48\pi \widetilde{\mu}_\theta} \Big), \\
\hat{\lambda} =& \widetilde{\lambda}_1 - \frac{1}{16\pi}\frac{1}{\widetilde{\mu}_\theta}\Big(2 \widetilde{\lambda}^2_3 + 2 \widetilde{\lambda}_3\widetilde{\lambda}_4 + \widetilde{\lambda}^2_4 + |\widetilde{\lambda}_5|^2 -48 \RE(\widetilde{\lambda}_6\widetilde{\lambda}_7 ) + 48 |\widetilde{\lambda}_6|^2 \Big), \\
\hat{\mu}_3^2 =& \widetilde{\mu}^2_\phi - \frac{\widetilde{\mu}_\theta}{4\pi}\Big(2 \widetilde{\lambda}_3 + \widetilde{\lambda}_4\Big) +\frac{1}{16\pi^2} \bigg(\frac{1}{8}(3\bar{g}^2_3 + \bar{g}'^2_3 )(2\widetilde{\lambda}_3 + \widetilde{\lambda}_4) - \widetilde{\lambda}^2_3 - \widetilde{\lambda}_3 \widetilde{\lambda}_4 -\widetilde{\lambda}^2_4 \nonumber \\
&  \quad + 3 \widetilde{\lambda}_2 (2\widetilde{\lambda}_3+\widetilde{\lambda}_4) + 18 \RE(\widetilde{\lambda}_7 \widetilde{\lambda}_6) -3|\widetilde{\lambda}_5|^2 \Big(\ln\Big(\frac{\Lambda_{3}'}{2 \widetilde{\mu}_\theta}\Big) + \frac{1}{2} \Big)-3|\widetilde{\lambda}_7|^2 \Big(\ln\Big(\frac{\Lambda_{3}'}{3 \widetilde{\mu}_\theta}\Big) + 2 \Big) \nonumber \\
& - \quad 9|\widetilde{\lambda}_6|^2 \Big( \ln\Big(\frac{\Lambda_{3}'}{ \widetilde{\mu}_\theta}\Big) + \frac{1}{2} \Big) + \frac{1}{8} \Big(-3\bar{g}^4_3 - \bar{g}'^4_3 + 4(3\bar{g}^2_3 + \bar{g}'^2_3)(2\widetilde{\lambda}_3 + \widetilde{\lambda}_4) \nonumber \\
& \quad - 16 (\widetilde{\lambda}^2_3 + \widetilde{\lambda}_3
\widetilde{\lambda}_4 + \widetilde{\lambda}^2_4 ) \Big) \ln \Big( \frac{\Lambda_{3}'}{2 \widetilde{\mu}_\theta} \Big)   \bigg).
\end{align}
Logarithmic terms could again be replaced by the exact RG evolution in the final effective theory (\ref{eq:final_3d_lag}). In order to make a connection to the existing lattice results of Ref.~\cite{Kajantie:1995kf}, we fix the renormalization scale of the final effective theory as $\Lambda_{3}'' = \hat{g}^2_3$.

\subsection{Effects from 6-dim. operators}
\label{sec:DR_validity}

In our $O(g^4)$ DR procedure, we omitted the effects coming from higher-order operators. For the validity of DR, it is important to estimate the effect of these neglected operators; indeed, we observe regions in the parameter space where the parameter $x$ becomes negative, signaling the breakdown of the last step of DR (see Fig. 2 in the companion paper \cite{Andersen:2017ika}). In this section we discuss a few simple 6-dim. operators and give perturbative estimates for the validity of our effective theories, yet we emphasize that a full evaluation of all 6-dim. operators for a more comprehensive estimate is a formidable task.

\subsubsection{6-dim. operators from the first DR step}

By using the effective potential and the background field method (see Appendix~\ref{sec:effectivepot}), it is trivial to obtain the three-dimensional coefficients for the most simple 6-dim. operators in the heavy-scale effective theory, of which we analyze $(\phi_1^{\dagger}\phi_1)^3$ and $(\phi_2^{\dagger}\phi_2)^3$. The magnitude of these correlators provide a rough estimate of the validity of the first step of DR.

Coefficients of the aforementioned operators in the effective potential read
\begin{align}
V_{6,1} &= \frac{\zeta(3)}{3(4\pi)^4T^2}\left[ 30\lambda_1^3 + \frac{1}{4}\lambda_3^3+ \lambda_+^3+ \lambda_-^3 + \frac{3}{32}g^6 + \frac{3}{64}(g^2+g'^2)^3  \right], \\
V_{6,2} &= \frac{\zeta(3)}{3(4\pi)^4T^2}\left[ 30\lambda_2^3 + \frac{1}{4}\lambda_3^3+ \lambda_+^3+ \lambda_-^3 + \frac{3}{32}g^6 + \frac{3}{64}(g^2+g'^2)^3 - \frac{21}{2}g_Y^6\right],
\end{align}
where $\lambda_\pm \equiv \frac12 (\lambda_3 + \lambda_4 \pm \lambda_5)$. In the SM case in Ref.~\cite{Kajantie:1995dw}, in Eq. (201) it was shown that the dominant 6-dim. contribution comes from the top quark (compared to the Higgs self-coupling and gauge contributions) and that the relative shift caused by this 6-dim. operator to the vev of the Higgs in the effective theory is of the order of one percent. Therefore, we get a rough estimate of the validity of DR by investigating the ratios 
\begin{align}
\frac{V^{\text{BSM}}_{6,1}}{V^{\text{SM}}_{6,2}} \quad \quad \text{and} \quad \quad
\frac{V^{\text{BSM}}_{6,2}}{V^{\text{SM}}_{6,2}}.
\end{align}
If these ratios are large, we can expect that even the first step of DR fails. These estimates have been included in the parameter-space scans of the companion paper \cite{Andersen:2017ika}.

\subsubsection{Validity of the SM-like effective theory}
\label{sec:dim6-step3}

The phase-transition analysis presented in the companion paper~\cite{Andersen:2017ika} is based on the assumption that dynamics of the transition can effectively be described by the one-doublet theory discussed in Section \ref{sec:SM-like}. We previously argued that this approximation can be justified in the presence of a sizable mixing term $\mu^2_{12} \phi^\dagger_1 \phi_2$, but shall now study the reliability of the last DR step in more detail by explicitly including the operator $(\phi^\dagger \phi)^3$ in the final effective theory and studying it perturbatively.  

The scalar potential in this theory reads
\begin{align}
V(\phi) = \hat{\mu}_3^2 \phi^\dagger \phi + \hat{\lambda}_3 (\phi^\dagger \phi)^2 + \hat{\Lambda}_6 (\phi^\dagger \phi)^3,
\end{align}
where the coefficients $\hat{\mu}_3^2,\hat{\lambda}_3$ are as in Section~\ref{sec:heavy_matching} and $\hat{\Lambda}_6$ is to be matched by computing the six-point function in the diagonalized theory of Eq.~(\ref{eq:3d_diagonal}). Due to the presence of $\widetilde{\lambda}_6$ and $\widetilde{\lambda}_7$ terms, the six-point function contains one-particle-reducible contributions that are not reproduced by the SM-like effective theory. In particular, the matching relation is dominated by a tree-level diagram proportional to $|\widetilde{\lambda}_6|^2$ \cite{Laine:1996ms}. Including only this leading-order contribution, we obtain a matching relation for the 6-dim. coefficient 
\begin{align}
\hat{\Lambda}_6 = \frac83 \frac{|{\widetilde{\lambda}^2_6}|}{m^2_\theta}.
\end{align} 

The effect of the operator $(\phi^\dagger \phi)^3$ can be probed by calculating the effective potential in this effective theory. Since the theory is purely spatial with all temperature dependence being nested in the parameters themselves, we face no issues regarding thermal resummation. For the SM-like effective theory without the 6-dim. operator, the two-loop effective potential has previously been obtained in Ref.~\cite{Farakos:1994kx}. Our calculation is similar to theirs but with modified couplings and masses (see Appendix \ref{sec:dim6-Veff}). Finally, let us point out that since the 6-point coupling $\hat{\Lambda}_6$ is matched already at tree level, there exist three-loop and higher diagrams that may be of numerical importance, but have been left out from our error estimate. 

If the minimum of the potential is shifted significantly by the inclusion of the 6-dim. operator, we can conclude that the SM-like effective theory is inadequate for describing the EWPT without accounting for higher-order operators, and basing the analysis on the lattice results of Ref.~\cite{Kajantie:1995kf} is not reliable. In the companion paper, two points\footnote{The exact input parameters used were $m_{A_0} = 270$ GeV ($x = 0.108$) and $m_{A_0} = 280$ GeV ($x = 0.063$), with $t_\beta = 2,\ m_{H_0}=180$ GeV, $\mu=75$ GeV, $m_{H^\pm} = m_{A_0}$ and $\cos(\beta-\alpha) = 0$ for both cases.} with a sizable mass splitting between the scalar states $A_0$ and $H_0$ were analyzed using the three-dimensional effective potential (see Fig.~3 in Ref.~\cite{Andersen:2017ika}). While near the crossover boundary ($x \approx 0.11$), the SM-like effective theory was found to work well, increasing $m_{A_0}$ caused effects from the 6-dim. operator to became large and the location of the degenerate minimum was shifted considerably. This is caused by the large portal couplings resulting from the mass hierarchy between the $H_0$ and $A_0$ eigenstates, which, after diagonalization, leads to a large ${\widetilde{\lambda}_6}$ in the 6-dim. coefficient. This is an unfortunate result, since in perturbation theory, very strong phase transitions are obtained precisely in the large-$m_{A_0}$ region \cite{Dorsch:2014qja,Basler:2016obg}. 

The conclusion is that although qualitative understanding of the phase structure of 2HDM can be obtained by integrating out the second doublet, accurate determination of equilibrium quantities relevant for applications is beyond the reach of the SM-like effective theory and clearly calls for new simulations with two dynamical doublets. However, the SM-like theory may still be used to accurately find the critical line where a crossover turns into a first-order transition, which is something not visible in perturbation theory.

\section{Discussion}
\label{sec:discussion}

In this work, we have derived three-dimensional high-temperature effective theories for the 2HDM. We emphasize that each of these theories is able to reproduce long-distance physics of the full 2HDM and can be studied on the lattice. An advantage of this three-dimensional approach combined with lattice simulations is the accurate treatment of the IR physics at high temperatures, which cannot be reliably described by purely perturbative methods. In particular, dimensional reduction naturally incorporates resummations of higher-order IR-sensitive contributions, and results obtained from the three-dimensional effective theories can hence be expected to be more reliable than those obtained with resummed effective potential alone. 

However, as the 2HDM is often studied with fairly large couplings in
the scalar sector, the validity of the perturbative expansion
used for dimensional reduction must be addressed in order to
reach reliable results. In practice, this requires a careful analysis
of higher-order corrections to the parameters of the effective theory,
and estimates for neglected higher-order operators. In the context of
a finite-$T$ effective potential, perturbativity of the one-loop
contribution must be analyzed by including full two-loop corrections
and investigating relative convergence. In the inert
    doublet model, a resummed two-loop effective potential has been
  calculated in Ref.~\cite{Laine:2017hdk} with considerably large
  differences in the results relative to a one-loop analysis.

Using existing lattice results, we have performed a non-perturbative analysis of the three-dimensional theory where the second doublet has been integrated out. In the alignment limit of the 2HDM, our results are presented in a companion paper~\cite{Andersen:2017ika}. We have also found that the analysis based on this SM-like effective theory is unreliable in the presence of large portal couplings. In the near future, we plan on extending our study to lattice simulations of two dynamical doublets, and non-perturbatively determining the thermodynamic quantities of interest, such as the latent heat and surface tension, in addition to the character and strength of the transition that were analyzed in the companion paper~\cite{Andersen:2017ika}. This would allow us to set a concrete benchmark for the accuracy of the widely-used perturbative treatment with a finite-$T$ effective potential.

\section*{Acknowledgments}

TT has been supported by the Vilho, Yrj\"{o} and Kalle
V\"{a}is\"{a}l\"{a} Foundation, as well as by the Swiss National Science
Foundation (SNF) under grant 200020-168988. TG, LN and TT have been
supported by the Academy of Finland grant no.\ 1303622,
as well as by the European Research Council grant
no.\ 725369. LN was also supported by the Academy
of Finland grant no.\ 308791. DJW (ORCID ID
0000-0001-6986-0517) was supported by the Academy of Finland grant
no.~286769 and by the Research Funds of the University of Helsinki. AH was in part funded by the Danish National Research Foundation (DNRF91). The authors would like to thank Jens Andersen, Tom\'a\v{s} Brauner, Stephan J. Huber, Kimmo Kainulainen, Keijo Kajantie, Venus Keus, Mikko Laine, Jose M. No, Hiren H. Patel, Michael J. Ramsey-Musolf, Kari Rummukainen, Anders Tranberg, Ville Vaskonen and Aleksi Vuorinen for discussions.


\appendix

\section{Diagonalization into a doublet-diagonal basis}
\label{sec:diagonalization}

A unitary transformation diagonalizing the 2HDM scalar potential can be written as
\begin{align}
\begin{pmatrix}
    \phi_1 \\
    \phi_2 \\
  \end{pmatrix}
\equiv
\begin{pmatrix}
    \alpha & \beta \\
    -\beta & \alpha^* \\
\end{pmatrix}
\begin{pmatrix}
    \phi \\
    \theta \\
  \end{pmatrix},
\end{align}
where
\begin{align}
\label{eq:diag_alpha}
\alpha =& \pm e^{i\varphi} \sqrt{\frac{\mu^2_{22} - \mu^2_{11} + \bar{\Omega}}{2\bar{\Omega}}}, \\
\beta =& \sqrt{\frac{\mu^2_{11} - \mu^2_{22} + \bar{\Omega}}{2\bar{\Omega}}}, \\ 
\varphi =& \tan^{-1} \frac{\IM(\mu^2_{12})}{\RE(\mu^2_{12})}, \\	
\bar{\Omega} =& \; \sqrt{(\mu^2_{11} - \mu^2_{22})^2 + 4 |\mu^{2}_{12}|^2}.
\end{align}
The sign in Eq.~(\ref{eq:diag_alpha}) should be chosen to match that of $\RE \, \mu^2_{12}$ (for $\RE \, \mu^2_{12} = 0$, the sign is determined by that of $\IM \, \mu^2_{12}$, and we choose $\varphi = \pi/2$). This transformation generalizes the rotation of Eqs.~(6.16-17) of Ref.~\cite{Laine:1996ms} to complex $\mu^{2}_{12}$.
Mass parameters in the diagonal basis read
\begin{align}
\mu^2_\phi = \frac{1}{2}( \mu^2_{11} + \mu^2_{22} - \bar{\Omega}),  \\
\mu^2_\theta = \frac{1}{2}( \mu^2_{11} + \mu^2_{22} + \bar{\Omega}).
\end{align}

When the mixing mass is considerably heavy, $\mu_{12} \gtrsim gT$, there will generally be a mass hierarchy between the eigenstates $\phi, \theta$. When diagonalization is applied to the effective theory of Eq.~(\ref{eq:lag_3d_scalars}), we may integrate out the heavier doublet $\theta$ in the limit of a large mixing term to obtain the SM-like effective theory described in Section \ref{sec:SM-like}. See the companion paper \cite{Andersen:2017ika} for an application of this approach. However, in the case of small $\bar{\Omega}$, both doublets can be light and must be dynamically included in the three-dimensional lattice simulations.

Scalar self-couplings in diagonal basis are given by
\begin{align}
  \begin{pmatrix}
    \widetilde{\lambda}_1 \\
    \widetilde{\lambda}_2 \\
    \widetilde{\lambda}_3 \\
    \widetilde{\lambda}_4 \\
    \widetilde{\lambda}_5 /2 \\
    \widetilde{\lambda}_6 \\
    \widetilde{\lambda}_7
  \end{pmatrix}
=
M
\begin{pmatrix}
    \bar{\lambda}_{1,3} \\
    \bar{\lambda}_{2,3} \\
    \bar{\lambda}_{3,3} \\
    \bar{\lambda}_{4,3} \\
    \bar{\lambda}_{5,3} \\
    \bar{\lambda}^*_{5,3} \\
    \bar{\lambda}_{6,3} \\
    \bar{\lambda}^*_{6,3} \\
    \bar{\lambda}_{7,3} \\
    \bar{\lambda}^*_{7,3}
  \end{pmatrix},
\end{align}
where

    \begin{equation}
  \resizebox{\textwidth}{!}{$%
M=\left(
\begin{array}{cccccccccc}
 \alpha ^2 \left(\alpha ^*\right)^2 & \beta ^4 & \alpha  \beta ^2 \alpha ^* & \alpha  \beta ^2
   \alpha ^* & \frac{1}{2} \beta ^2 \left(\alpha ^*\right)^2 & \frac{\alpha ^2 \beta ^2}{2} &
   -\alpha  \beta  \left(\alpha ^*\right)^2 & -\alpha ^2 \beta  \alpha ^* & -\alpha  \beta ^3
   & -\beta ^3 \alpha ^* \\
 \beta ^4 & \alpha ^2 \left(\alpha ^*\right)^2 & \alpha  \beta ^2 \alpha ^* & \alpha  \beta ^2
   \alpha ^* & \frac{1}{2} \beta ^2 \left(\alpha ^*\right)^2 & \frac{\alpha ^2 \beta ^2}{2} &
   \beta ^3 \alpha ^* & \alpha  \beta ^3 & \alpha ^2 \beta  \alpha ^* & \alpha  \beta 
   \left(\alpha ^*\right)^2 \\
 2 \alpha  \beta ^2 \alpha ^* & 2 \alpha  \beta ^2 \alpha ^* & \beta ^4+\alpha ^2 \left(\alpha
   ^*\right)^2 & -2 \alpha  \beta ^2 \alpha ^* & -\beta ^2 \left(\alpha ^*\right)^2 & -\alpha
   ^2 \beta ^2 & \alpha  \beta  \left(\alpha ^*\right)^2-\beta ^3 \alpha ^* & \alpha  \beta 
   \left(\alpha  \alpha ^*-\beta ^2\right) & \alpha  \beta ^3-\alpha ^2 \beta  \alpha ^* &
   \beta ^3 \alpha ^*-\alpha  \beta  \left(\alpha ^*\right)^2 \\
 2 \alpha  \beta ^2 \alpha ^* & 2 \alpha  \beta ^2 \alpha ^* & -2 \alpha  \beta ^2 \alpha ^* &
   \beta ^4+\alpha ^2 \left(\alpha ^*\right)^2 & -\beta ^2 \left(\alpha ^*\right)^2 & -\alpha
   ^2 \beta ^2 & \alpha  \beta  \left(\alpha ^*\right)^2-\beta ^3 \alpha ^* & \alpha  \beta 
   \left(\alpha  \alpha ^*-\beta ^2\right) & \alpha  \beta ^3-\alpha ^2 \beta  \alpha ^* &
   \beta ^3 \alpha ^*-\alpha  \beta  \left(\alpha ^*\right)^2 \\
 \beta ^2 \left(\alpha ^*\right)^2 & \beta ^2 \left(\alpha ^*\right)^2 & -\beta ^2
   \left(\alpha ^*\right)^2 & -\beta ^2 \left(\alpha ^*\right)^2 & \frac{\left(\alpha
   ^*\right)^4}{2} & \frac{\beta ^4}{2} & \beta  \left(\alpha ^*\right)^3 & -\beta ^3 \alpha
   ^* & \beta ^3 \alpha ^* & -\beta  \left(\alpha ^*\right)^3 \\
 2 \alpha  \beta  \left(\alpha ^*\right)^2 & -2 \beta ^3 \alpha ^* & \beta ^3 \alpha ^*-\alpha
    \beta  \left(\alpha ^*\right)^2 & \beta ^3 \alpha ^*-\alpha  \beta  \left(\alpha
   ^*\right)^2 & -\beta  \left(\alpha ^*\right)^3 & \alpha  \beta ^3 & \alpha  \left(\alpha
   ^*\right)^3-\beta ^2 \left(\alpha ^*\right)^2 & -2 \alpha  \beta ^2 \alpha ^* & \alpha 
   \beta ^2 \alpha ^*-\beta ^4 & 2 \beta ^2 \left(\alpha ^*\right)^2 \\
 2 \alpha  \beta ^3 & -2 \alpha ^2 \beta  \alpha ^* & \alpha ^2 \beta  \alpha ^*-\alpha  \beta
   ^3 & \alpha ^2 \beta  \alpha ^*-\alpha  \beta ^3 & -\beta ^3 \alpha ^* & \alpha ^3 \beta  &
   \alpha  \beta ^2 \alpha ^*-\beta ^4 & 2 \alpha ^2 \beta ^2 & \alpha ^3 \alpha ^*-\alpha ^2
   \beta ^2 & -2 \alpha  \beta ^2 \alpha ^* \\
\end{array}
\right).
    $}
    \end{equation}

\section{Parameterization of the scalar sector}
\label{sec:parameter_constraints}

\subsection{Stability and unitarity constraints}

For the potential defined in Eq.~(\ref{eq:scalar_potential}), boundedness from below is achieved if 
\begin{equation}
\lambda_1 > 0, \quad \lambda_2 > 0, \quad \lambda_3 > - 2 \sqrt{\lambda_1 \lambda_2} \quad \text{and} \quad \lambda_3 + \lambda_4 - |\lambda_5| > -2 \sqrt{\lambda_1 \lambda_2}.
\end{equation}
These have been obtained in Refs. \cite{Deshpande:1977rw,Klimenko:1984qx,Ginzburg:2005yw,Kastening:1992by,Gunion:2002zf,Haber:1994mt}. Furthermore, to guarantee that the $v=246$ GeV vacuum is a global minimum of the potential, we impose the following constraint (in the case of a softly-broken $Z_2$ symmetry): 
\begin{equation}
\bigg[ \bigg( \frac{m_{H^{\pm}}^2}{v^2} + \frac{\lambda_4}{2}\bigg)^2 - \frac{|\lambda_5|^2}{4}\bigg]\bigg[ \frac{m_{H^{\pm}}^2}{v^2} + \frac{2\sqrt{\lambda_1 \lambda_2}-\lambda_3}{2} \bigg] > 0 .
\end{equation}
This relation, as well as discussion on the possibility of a metastable vacuum in the 2HDM, can be found in Refs. \cite{Barroso:2013awa,Ivanov:2015nea}.

Requirement of unitarity conservation, discussed in Refs.~\cite{Akeroyd:2000wc,Ginzburg:2003fe,Ginzburg:2005dt}, further limits the possible values of the $\lambda$'s. Tree-level unitarity constraints can be written in the form
\begin{align}
| \Lambda^{Z_{2}}_{Y \sigma \pm} | &< 8 \pi,
\end{align}
where in our parameterization
\begin{align}
    \Lambda^{\text{even}}_{21\pm} &=  \lambda_{1} + \lambda_{2} \pm \sqrt{(\lambda_{1}-\lambda_{2})^{2} +  | \lambda_{5} |^{2}},\\
    \Lambda^{\text{even}}_{01\pm} &=   \lambda_{1} + \lambda_{2} \pm \sqrt{(\lambda_{1}-\lambda_{2})^{2} +  \lambda_{4}^{2}} ,\\
    \Lambda^{\text{even}}_{00\pm} &=  3 (\lambda_{1} + \lambda_{2}) \pm \sqrt{9(\lambda_{1}-\lambda_{2})^{2} + (2 \lambda_{3} + \lambda_{4})^{2}} ,\\
    \Lambda^{\text{odd}}_{21} &= \lambda_{3}+\lambda_{4} ,\\
    \Lambda^{\text{odd}}_{20} &= \lambda_{3}-\lambda_{4} ,\\
    \Lambda^{\text{odd}}_{01\pm} &= \lambda_{3} \pm | \lambda_{5} | ,\\
    \Lambda^{\text{odd}}_{00\pm} &= \lambda_{3} + 2 \lambda_{4} \pm 3 |\lambda_{5}| .
\end{align}

\subsection{Parameters of the four-dimensional theory in terms of physical quantities}
\label{sec:params_tree}

Here we have set $\lambda_5$ and $\mu_{12}^2$ to be real and defined $\Omega^2 \equiv m_{H_0}^2-\mu^2(t_\beta + t_\beta^{-1})$, where $\mu^2 \equiv -\RE \, \mu_{12}^2$. Mass parameters and couplings of the scalar sector are given by 
\begin{align}
\mu _{11}^2 &= \mu^2 t_\beta - \frac{1}{2} [m_h^2 + (m_{H_0}^2 - m_h^2) c_{\beta-\alpha} (c_{\beta-\alpha}+s_{\beta-\alpha} t_\beta) ] ,\\ 
\mu_{22}^2 &= \mu^2 t_\beta^{-1} - \frac{1}{2} [m_h^2 + (m_{H_0}^2 - m_h^2) c_{\beta-\alpha} (c_{\beta-\alpha}-s_{\beta-\alpha} t_\beta^{-1}) ] ,\\
v^2 \lambda_1 &= \frac{1}{2}\bigg\{ m_h^2 + \Omega^2 t_\beta^2 - (m_{H_0}^2 - m_h^2)[1-(s_{\beta-\alpha}+c_{\beta-\alpha} t_\beta^{-1})^2] t_\beta^2\bigg\} ,\\
v^2 \lambda_2 &= \frac{1}{2}\bigg\{ m_h^2 + \Omega^2 t_\beta^{-2} - (m_{H_0}^2 - m_h^2)[1-(s_{\beta-\alpha}-c_{\beta-\alpha} t_\beta)^2] t_\beta^{-2} \bigg\} ,\\
v^2 \lambda_3 &= 2 m_{H^{\pm}} + \Omega^2 -m_h^2 - (m_{H_0}^2 - m_h^2)[1+(s_{\beta-\alpha}+c_{\beta-\alpha} t_\beta^{-1})(s_{\beta-\alpha}-c_{\beta-\alpha} t_\beta)] ,\\
v^2 \lambda_4 &= m_{A_0}^2 - 2 m_{H^{\pm}} + m_{H_0}^2 - \Omega^2 ,\\
v^2 \lambda_5 &= m_{H_0}^2 - m_{A_0}^2 - \Omega^2.
\end{align}
These can also be found in Ref.~\cite{Dorsch:2017nza}, but note that there are misprints in powers of $t_\beta$ in the equations for $\lambda_1$ and $\lambda_2$. 

\section{Details of dimensional reduction}

In principle, all zero-momentum correlator functions necessary for the matching relations could be read from the effective potential.  While for pure scalars this is straightforward, calculating the effective potential for mixed scalar-gauge correlators is more subtle. Therefore, we have chosen to evaluate the gauge correlators by a direct diagram-by-diagram calculation, in analogy to Refs.~\cite{Kajantie:1995dw,Brauner:2016fla}. For pure-scalar correlators at one-loop level, we apply both of these methods as a cross-check of the correctness of our calculation.

At two-loop level, we only need doublet self energies, which we have calculated diagrammatically rather than evaluating the two-loop effective potential as was done in Ref.~\cite{Kajantie:1995dw} in the case of the SM. For this calculation, one has to evaluate a decent amount of individual diagrams; we choose not to present intermediate results in a diagram-by-diagram form, and only the final results have been given in Section \ref{sec:matching}. However, in Appendix \ref{sec:integrals} we provide a list of the required integrals for this calculation.

\subsection{Matching relations from the effective potential}
\label{sec:effectivepot}
An economic way of calculating the scalar correlators needed for dimensional reduction is to use the effective potential; we illustrate the procedure in this appendix (see also \cite{Helset:2017ab}). In order to calculate the effective potential, one decomposes the 
scalar fields into quantum and classical fields, $\phi_i \rightarrow \phi_i + \varphi_i$. The
quantum fields $\phi_i$ are integrated over in the path-integral formalism in the usual way, while
the classical fields $\varphi_i$ are not. The functional form of the effective potential can be found by expanding the resulting potential in the background fields.

Following Ref.~\cite{Kajantie:1995dw}, we include one-loop contributions  
from scalars, gauge bosons and fermions to the effective potential:
\begin{equation}
V^{1\text{-loop}}_{\text{eff}} = \mathcal{C}_S(m) + \mathcal{C}_V(M) + \mathcal{C}_F(m_f),
\end{equation}
where
\begin{align}
\label{eq:Cs}
\mathcal{C}_S(m) &\equiv -\sumint K \log\left(\frac{1}{K^2+m^2}\right)^{1/2} = J_b(m) \\
\label{eq:Cv}
\mathcal{C}_V(M) &\equiv -\sumint K \log\left(\det\frac{\delta_{\mu\nu}-K_{\mu}K_{\nu}/K^2}
{K^2+M^2}\right)^{1/2}=(3-2\epsilon)J_b(M) \\
\label{eq:Cf}
\mathcal{C}_F(m_f) &\equiv -\sumint {\{K\}} \log\left(\frac{1}{i\slashed K+m_f}\right)^{1/2}	
=-4J_f(m_f),
\end{align}
where $m$, $M$, and $m_f$ are the mass eigenvalues for the scalars, gauge bosons, and fermions, 
respectively. In Eq.~(\ref{eq:Cv}) the prefactor comes from the trace of the projection
operator in the gauge field propagator, in $d=3-2\epsilon$ dimensions. Fermions in consideration are Dirac fermions, as indicated by the prefactor in Eq.~(\ref{eq:Cf}).
The integrals $J_b(m)$ and $J_f(m_f)$ are given in Eqs.~(\ref{eq:Jb}) and (\ref{eq:Jf}). Note that for one-loop matching, we only need the non-zero mode contributions of these sum-integrals.

We can extract the mass matrix from the quadratic parts in the gauge, scalar, and fermion fields. This matrix can be diagonalized and, using different choices of background fields, we extract the contributions to
the correlators at zero external momentum. Due to our power counting scheme, at one-loop level only the $\phi^\dagger_1 \phi_2$ correlator and its Hermitian conjugate are affected by the mass-mixing term. For all other correlation functions, the mass-mixing effects occur only at a higher order. Therefore, in order to keep our illustration simple, in this section we only consider the $Z_2$ symmetric case $\lambda_6=\lambda_7=0$ and $\mu^2_{12}=0$, and we set $\text{Im}(\lambda_5)=0$ for simplicity. The effective potential can be expanded in the background fields,
\begin{align}
	\label{eq:scalarexpansion}
	V_{\text{eff}} = V_{11}\varphi_1^{\dagger}\varphi_1 +  V_{22}\varphi_2^{\dagger}\varphi_2 
		+ V_{1}\left[\varphi_1^{\dagger}\varphi_1\right]^2
		+ V_{2}\left[\varphi_2^{\dagger}\varphi_2\right]^2
		+ V_{3}\left[\varphi_1^{\dagger}\varphi_1\right]\left[\varphi_2^{\dagger}\varphi_2\right]
		\nonumber\\
		+ V_{4}\left[\varphi_1^{\dagger}\varphi_2\right]\left[\varphi_2^{\dagger}\varphi_1\right]
		+ \frac{V_{5}}{2}\left[\left[\varphi_1^{\dagger}\varphi_2\right]^2
		+ \left[\varphi_2^{\dagger}\varphi_1\right]^2\right].
\end{align}
By determining the above coefficients $V$, 
we can find the correlators needed for dimensional reduction.
We shift the scalar fields $\phi_i \rightarrow \phi_i + \varphi_i$, and focus first on the bilinear
scalar terms, 

\begin{align}
	V_{\textrm{scalar}}[\phi_1 + \varphi_1, \phi_2 + \varphi_2] 
	={}& -\frac{1}{2}\tilde{m}_{11}^2\phi_1^{\dagger}\phi_1 
	- \frac{1}{2}\tilde{m}_{22}^2\phi_2^{\dagger}\phi_2 \nonumber  \\
	&+ \lambda_1 \Big[ \phi_1^{\dagger}\varphi_1 + \varphi_1^{\dagger}\phi_1\Big]^2 
	+ \lambda_2 \Big[ \phi_2^{\dagger}\varphi_2 + \varphi_2^{\dagger}\phi_2\Big]^2  \nonumber \\
	&+ \lambda_3\Big[\phi_1^{\dagger}\varphi_1 + \varphi_1^{\dagger}\phi_1\Big]
	\Big[ \phi_2^{\dagger}\varphi_2	+ \varphi_2^{\dagger}\phi_2\Big] \nonumber  \\ 
	&+ \lambda_4\Big[ (\phi_1^{\dagger}\phi_2)(\varphi_2^{\dagger}\varphi_1)
		+ (\phi_2^{\dagger}\phi_1)(\varphi_1^{\dagger}\varphi_2) 
		\nonumber \\
		&+ (\phi_1^{\dagger}\varphi_2 + \varphi_1^{\dagger}\phi_2)(\phi_2^{\dagger}\varphi_1 
	+ \varphi_2^{\dagger}\phi_1)\Big]  \nonumber \\
	&+ \frac{1}{2}\lambda_5 \Big[ 2(\phi_1^{\dagger}\phi_2)(\varphi_1^{\dagger}\varphi_2)
		+ 2(\phi_2^{\dagger}\phi_1)(\varphi_2^{\dagger}\varphi_1) 
		\nonumber \\
		&+ \big[ \phi_1^{\dagger}\varphi_2 + \varphi_1^{\dagger}\phi_2\big]^2 
		+ \big[ \phi_2^{\dagger}\varphi_1 + \varphi_2^{\dagger}\phi_1\big]^2\Big] 
		+ \mathcal{O}(\phi_i^3) + \mathcal{O}(\phi_i^4)
\end{align}
where $\tilde{m}_{11}^2 = -2\mu_{11}^2 - 4\lambda_1 \varphi_1^{\dagger}\varphi_1 
- 2\lambda_3\varphi_2^{\dagger}\varphi_2$ 
and  $\tilde{m}_{22}^2 = -2\mu_{22}^2 - 4\lambda_2 \varphi_2^{\dagger}\varphi_2 
- 2\lambda_3\varphi_1^{\dagger}\varphi_1$. 

We encounter a complication when trying to distinguish the
$V_3,V_4$ and $V_5$ contributions. In order to separate them, we make three different choices for the background fields
\begin{align}
	\label{eq:case1}
	\textrm{Case } 1: \quad \varphi_1 &= \frac{1}{\sqrt{2}} 
	\begin{pmatrix} 0 \\ v_1 \end{pmatrix}, \quad \varphi_2 = \frac{1}{\sqrt{2}} 
	\begin{pmatrix} 0 \\ v_2 \end{pmatrix} \\
	\label{eq:case2}
	\textrm{Case } 2: \quad \varphi_1 &= \frac{1}{\sqrt{2}} 
	\begin{pmatrix} 0 \\  v_1 \end{pmatrix}, \quad \varphi_2 = \frac{1}{\sqrt{2}} 
	\begin{pmatrix} 0 \\  iw_0\end{pmatrix} \\
	\label{eq:case3}
	\textrm{Case } 3: \quad \varphi_1 &= \frac{1}{\sqrt{2}} 
	\begin{pmatrix} 0 \\ v_1 \end{pmatrix}, \quad \varphi_2 = \frac{1}{\sqrt{2}} 
	\begin{pmatrix} w_+ \\  0 \end{pmatrix}.
\end{align}
For each case, we diagonalize the mass matrix, and evaluate the scalar part of the effective potential by using background-field-dependent mass-squared eigenvalues, and the sum-integral of Eq.~(\ref{eq:Jb}). By expanding to order $O(g^2)$ in mass parameters and to $O(g^4)$ in couplings, we find expressions of the form   
\begin{align}
\label{eq:case1series}
\text{Case 1:}& \quad V_{\textrm{eff}} =\frac{1}{2} V_{11}v_1^2 + \frac{1}{2}V_{22}v_2^2 +\frac{1}{4} V_{1}v_1^4 
+\frac{1}{4} V_{2}v_2^4 + \frac{1}{4}(V_{3} + V_4 + V_5)v_1^2v_2^2, \\
\text{Case 2:}& \quad  V_{\textrm{eff}} = \frac{1}{2}V_{11}v_1^2 +\frac{1}{2} V_{22}w_0^2 +\frac{1}{4} V_{1}v_1^4  + \frac{1}{4}V_{2}w_0^4 +  \frac{1}{4}(V_{3}+ V_4 - V_5)v_1^2w_0^2, \\
\text{Case 3:}& \quad  V_{\textrm{eff}} = \frac{1}{2}V_{11}v_1^2 + \frac{1}{2}V_{22}w_+^2 +\frac{1}{4} V_{1}v_1^4 
+ \frac{1}{4}V_{2}w_+^4 + 
\frac{1}{4}V_{3}v_1^2w_+^2. 
\end{align}
We immediately obtain the coefficients $V_{11}, V_{22}, V_1$ and $V_2$ from the above expansions, and the remaining $V_3, V_4$ and $V_5$ can be solved from the linear system of coefficients of $v^2_1 v^2_2$, $v^2_1 \omega^2_0$ and $v^2_1 \omega^2_+$.

In the gauge sector, the covariant derivative $D_{\mu}\phi_i^{\dagger}D_{\mu}\phi_i$, where $i=1,2$,
couples the scalar and gauge fields. When the scalar fields are shifted by background fields, we get the bilinear terms
\begin{align}
	D_{\mu}\phi_1^{\dagger}D_{\mu}\phi_1 + D_{\mu}\phi_2^{\dagger}D_{\mu}\phi_2 
	\rightarrow & D_{\mu}\phi_1^{\dagger}D_{\mu}\phi_1 + D_{\mu}\phi_2^{\dagger} 
	D_{\mu}\phi_2 \nonumber \\
	&+ \frac{ig}{2} \vec{A}_{\mu} \Big[\varphi_1^{\dagger} \vec{\sigma} 
	\partial_{\mu} \phi_1 - \partial_{\mu} \phi_1^{\dagger} \vec{\sigma}\varphi_1 
	+ \varphi_2^{\dagger} \vec{\sigma} \partial_{\mu} \phi_2 - \partial_{\mu} 
	\phi_2^{\dagger} \vec{\sigma}\varphi_2 \Big] \nonumber \\
	&+ \frac{ig'}{2}B_{\mu}\Big[ \varphi_1^{\dagger}\partial_{\mu}\phi_1 
		- \partial_{\mu}\phi_1^{\dagger} \varphi_1 
		 + \varphi_2^{\dagger}\partial_{\mu}\phi_2 
	 - \partial_{\mu}\phi_2^{\dagger} \varphi_2 \Big] \nonumber \\
	 &+ \frac{1}{4}(\varphi_1^{\dagger}\varphi_1 + \varphi_2^{\dagger}\varphi_2)\Big[ 
	 g^2\vec{A}_{\mu}\vec{A}_{\mu} + g'^2B_{\mu}B_{\mu}\Big] \nonumber \\ 
	 &+ \frac{1}{2}gg'B_{\mu}\vec{A}_{\mu}\Big[ \varphi_1^{\dagger}\vec{\sigma}\varphi_1 
	 + \varphi_2^{\dagger}\vec{\sigma}\varphi_2 \Big].
\end{align}
The bilinear mixing terms between gauge bosons and Nambu-Goldstone bosons are removed by fixing the gauge
using the usual Faddeev-Popov gauge-fixing procedure. One must also include ghost fields, with
ghost masses and new ghost interactions proportional to the gauge fixing parameter $\xi$. However,
in the Landau gauge, $\xi=0$, the ghost masses and interactions vanish. As there is no
bilinear mixing between gauge bosons and ghost, we can safely go to the Landau gauge in the
ghost sector. The part bilinear in the gauge fields is given by the last two terms above.

We follow the same procedure as outlined above, making use of the three choices
of background fields. In Case 1, we get two massive charged, one massive neutral and one massless gauge boson with squared mass eigenvalues
\begin{align}
M_W^2 &= \frac{1}{4}g^2(v_1^2 + v_2^2), \\
M_Z^2 &= \frac{1}{4}(g^2 + g'^2)(v_1^2 + v_2^2).
\end{align}
In Case 2, we similarly find
\begin{align}
	M_W^2 &= \frac{1}{4}g^2(v_1^2 + w_0^2), \\
	M_Z^2 &= \frac{1}{4}(g^2 + g'^2)(v_1^2 + w_0^2),
\end{align}
and in finally Case 3, we find four massive gauge bosons, with eigenvalues
\begin{align}
M_W^2 &= \frac{1}{4}g^2(v_1^2 + w_+^2), \\
M_{\pm}^2 &= \frac{1}{8}\Big[ (g^2 + g'^2)(v_1^2 + w_+^2) \pm
\sqrt{(g^2-g'^2)^2(v_1^2 + w_+^2)^2 + 4g^2g'^2(v_1^2 - w_+^2)^2} \Big].
\end{align}
By using Eq.~(\ref{eq:Cv}) we can evaluate the coefficients of background fields in each case. Finally, we include contributions from the top quark,
that only couples to $\phi_2$, and thus only affects $V_{22}$ and $V_2$.

By collecting all contributions from the scalars, gauge bosons, and top quark at one-loop, coefficients of the effective potential of Eq.~(\ref{eq:scalarexpansion}) take the form
\begin{subequations}
\begin{align}
	\label{eq:effpotmass1}
	V_{11} =& \mu_{11}^2 + \frac{T^2}{12}\Big[ \frac{9}{4}g^2 + \frac{3}{4}g'^2 + 6\lambda_1 + 2\lambda_3 	+ \lambda_4
	\Big], \\
	\label{eq:effpotmass2}	
	V_{22} =& \mu_{22}^2 + \frac{T^2}{12}\Big[ \frac{9}{4}g^2 + \frac{3}{4}g'^2 + 6\lambda_2 + 2\lambda_3
 + \lambda_4+3g_{Y}^2 \Big], \\
	\label{eq:effpotint1}
	V_{1} =& \lambda_1 -\frac{1}{16(4\pi)^2}\Big[\frac{3}{\epsilon} + 3L_b - 2\Big]
	\Big(3g^4 + g'^4 + 2g^2g'^2\Big) \nonumber\\&- \frac{1}{(4\pi)^2}\Big(\frac{1}{\epsilon} + L_b\Big)
	\Big[12\lambda_1^2 + \frac{1}{2}\lambda_3^2 + \lambda_+^2 + \lambda_-^2\Big]  
	,\\
	\label{eq:effpotint2}	
	V_{2} =& \lambda_2 -\frac{1}{16(4\pi)^2}\Big[\frac{3}{\epsilon} + 3L_b - 2\Big]
	\Big(3g^4 + g'^4 + 2g^2g'^2\Big) \nonumber \\
	&- \frac{1}{(4\pi)^2}\Big(\frac{1}{\epsilon} + L_b\Big)
	\Big[12\lambda_2^2 + \frac{1}{2}\lambda_3^2 + \lambda_+^2 + \lambda_-^2\Big]
	+ \frac{3}{(4\pi)^2}\Big(\frac{1}{\epsilon} + L_f\Big)g_{Y}^4, \\
	\label{eq:effpotint3}
	V_{3} =& \lambda_3 -\frac{1}{8(4\pi)^2}\Big[\frac{3}{\epsilon} + 3L_b - 2\Big]
	\Big(3g^4 + g'^4 - 2g^2g'^2\Big) 
	\nonumber\\ &- \frac{1}{(4\pi)^2}\Big(\frac{1}{\epsilon} + L_b\Big)
	\Big[ 2(\lambda_1+\lambda_2)(3\lambda_3 + \lambda_4) + 2\lambda_3^2 + \lambda_4^2 
	+ \lambda_5^2 \Big],\\
	\label{eq:effpotint4}
	V_4 =& \lambda_4 -\frac{g^2g'^2}{2(4\pi)^2}\Big(\frac{3}{\epsilon} + 3L_b - 2\Big)
	 - \frac{1}{(4\pi)^2}\Big(\frac{1}{\epsilon} + L_b\Big)\Big[2(\lambda_1 + \lambda_2
	 +2\lambda_3 + \lambda_4)\lambda_4 + 4\lambda_5^2\Big], \\
	 \label{eq:effpotint5}
	 V_5 =& \lambda_5 -\frac{1}{(4\pi)^2}\Big(\frac{1}{\epsilon} + L_b\Big)\Big[2\lambda_5(\lambda_1 
	+ \lambda_2 + 2\lambda_3 + 3\lambda_4)\Big]. 
\end{align}
\end{subequations}
We have included both tree-level and one-loop contributions. From these coefficients, one can identify the required correlators and the corresponding counterterms. 

\subsection{Renormalization and one-loop $\beta$ functions}
\label{sec:4d_cts}

All fields and couplings appearing in the Lagrangian of Section \ref{sec:full4D} are the
renormalized ones, while the counterterms are included in $\delta\La$. We use the following conventions for the relations between the renormalized fields and couplings and their bare counterparts, denoted by the subscript $(b)$:
\begin{align}
\vec A_{\mu(b)} &\equiv Z^{1/2}_A \vec A_\mu = (1+\delta Z_A)^{1/2} \vec A_\mu,
&B_{\mu(b)} &\equiv Z^{1/2}_B B_\mu = (1+\delta Z_B)^{1/2} B_\mu, \nonumber \\
\phi_{1(b)} &\equiv Z^{1/2}_{\phi_1} \phi = (1+\delta Z_ {\phi_1})^{1/2} \phi_1, 
&\phi_{2(b)} &\equiv Z^{1/2}_{\phi_2} \phi = (1+\delta Z_ {\phi_2})^{1/2} \phi_2, \nonumber \\
q_{(b)} &\equiv Z^{1/2}_{q} q = (1+\delta Z_ {q})^{1/2} q, 
&t_{(b)} &\equiv Z^{1/2}_{t} t = (1+\delta Z_ {t})^{1/2} t,
\end{align}
for the fields, and
\begin{align}
\label{eq:bare_couplings} 
g_{(b)} &\equiv\Lambda^\epsilon(g + \delta g),
&g'_{(b)} &\equiv \Lambda^\epsilon(g' + \delta g'), \nonumber \\
g_{Y(b)} &\equiv Z^{-\frac{1}{2}}_{\phi_2} Z^{-\frac{1}{2}}_{q} Z^{-\frac{1}{2}}_{t} \Lambda^\epsilon(g_{Y} + \delta g_{Y}),
&\mu^2_{11(b)} &\equiv Z^{-1}_{\phi_1} (\mu^2_{11} + \delta \mu^2_{11}), \nonumber \\
\mu^2_{22(b)} &\equiv Z^{-1}_{\phi_2} (\mu^2_{22} + \delta \mu^2_{22}),
&\mu^2_{12(b)} &\equiv Z^{-\frac{1}{2}}_{\phi_1}Z^{-\frac{1}{2}}_{\phi_2} (\mu^2_{12} + \delta \mu^2_{12}), \nonumber \\
\lambda_{1(b)} &\equiv Z^{-2}_{\phi_1} \Lambda^{2\epsilon} (\lambda_1 + \delta \lambda_1), 
&\lambda_{2(b)} &\equiv Z^{-2}_{\phi_2} \Lambda^{2\epsilon} (\lambda_2 + \delta \lambda_2), \nonumber \\ 
\lambda_{3(b)} &\equiv Z^{-1}_{\phi_1}Z^{-1}_{\phi_2} \Lambda^{2\epsilon} (\lambda_3 + \delta \lambda_3), 
&\lambda_{4(b)} &\equiv Z^{-1}_{\phi_1}Z^{-1}_{\phi_2}  \Lambda^{2\epsilon} (\lambda_4 + \delta \lambda_4), \nonumber \\ 
\lambda_{5(b)} &\equiv Z^{-1}_{\phi_1}Z^{-1}_{\phi_2} \Lambda^{2\epsilon} (\lambda_5 + \delta \lambda_5),
&\lambda_{6(b)} &\equiv Z^{-\frac{3}{2}}_{\phi_1} Z^{-\frac{1}{2}}_{\phi_2} \Lambda^{2\epsilon} (\lambda_6 + \delta \lambda_6),  \nonumber \\ 
\lambda_{7(b)} &\equiv Z^{-\frac{1}{2}}_{\phi_1}Z^{-\frac{3}{2}}_{\phi_2} \Lambda^{2\epsilon} (\lambda_7 + \delta \lambda_7).
\end{align}
for the couplings, where $\Lambda$ is the renormalization scale.

In Landau gauge, the counterterms read explicitly 
\begin{align}
\delta Z_A =& \frac{g^2}{16\pi^2\epsilon}\Big(\frac{26-N_d}{6}-\frac{4}{3}N_f \Big),\\
\delta Z_B =&-\frac{g'^2}{96\pi^2\epsilon}\Big(N_d+N_f \bigl[2Y^2_\ell + Y^2_e + 3 ( 2Y^2_q + Y^2_u + Y^2_d)\bigr] \Big) = -\frac{g'^2}{96\pi^2\epsilon} \Big( N_d + \frac{40}{3}N_f \Big) ,\\
\delta Z_{\phi_n} =& \frac{1}{16\pi^2\epsilon}\Big(\frac{9}{4}g^2 + \frac{3}{4}g'^2 - 3 \delta_{2,n} g^2_{Y} \Big), \\
\delta Z_t =& \, 2 \delta Z_q = -\frac{g^2_Y}{16\pi^2\epsilon}, \\
\delta g =& -\frac{g^3}{16\pi^2\epsilon}\biggl(\frac{44-N_d}{12}-\frac{2}{3}N_f \biggr),\\
\delta g' =& \frac{g'^3}{192\pi^2\epsilon}\biggl(N_d+ \frac{40}{3}N_f \biggr),\\
\delta g_{Y} =& -\frac{g_{Y}}{16\pi^2\epsilon}\biggl( \frac{1}{3}g'^2 + 4 g^2_s \biggr), \\
\delta \mu^2_{11}\,=& \frac{1}{16\pi^2}\frac{1}{\epsilon}\Big(6 \lambda_1 \mu^2_{11} + (2\lambda_3 + \lambda_4)\mu^2_{22} + 6 \RE(\lambda_6 \mu^{2*}_{12}) \Big), \\
\delta \mu^2_{22}\,=& \frac{1}{16\pi^2}\frac{1}{\epsilon}\Big(6 \lambda_2 \mu^2_{22} + (2\lambda_3 + \lambda_4)\mu^2_{11} + 6 \RE(\lambda^*_7 \mu^{2*}_{12}) \Big), \\
\delta \mu^2_{12}\,=& \frac{1}{16\pi^2}\frac{1}{\epsilon}\Big(3 \lambda_6 \mu^2_{11} + 3 \lambda^*_7 \mu^2_{22} + 3 \lambda_5 \mu^{2*}_{12} + (2 \lambda_4 + \lambda_3) \mu^2_{12} \Big), \\
\delta \lambda_{1}\,=& \frac{1}{16\pi^2}\frac{1}{\epsilon}\frac{1}{2}\Big(24\lambda^2_1 + 2 \lambda^2_3 + 2\lambda_3 \lambda_4 + \lambda^2_4 + |\lambda_5|^2 + 12|\lambda_6|^2 \nonumber \\ 
& +\frac{3}{8}(3g^4 + {g'}^4 + 2 g^2{g'}^2) \Big), \quad  \\
\delta \lambda_{2}\,=& \frac{1}{16\pi^2}\frac{1}{\epsilon}\frac{1}{2}\Big(24\lambda^2_2 + 2 \lambda^2_3 + 2\lambda_3 \lambda_4 + \lambda^2_4 + |\lambda_5|^2 + 12|\lambda_7|^2 \nonumber \\ 
& +\frac{3}{8}(3g^4 + {g'}^4 + 2 g^2{g'}^2) - 6 g^4_{Y} \Big), \quad \\
\delta \lambda_{3}\,=& \frac{1}{16\pi^2}\frac{1}{\epsilon}\Big(2(\lambda_1 + \lambda_2)(3\lambda_3 + \lambda_4) + 2\lambda^2_3 +  \lambda^2_4 + |\lambda_5|^2 + \frac{3}{8}(3g^4 + {g'}^4 - 2 g^2{g'}^2) \nonumber \\ 
& + 2( |\lambda_6|^2+ |\lambda_7|^2) + 8 \RE(\lambda_6 \lambda_7) \Big), \quad  \\
\delta \lambda_{4}\,=& \frac{1}{16\pi^2}\frac{1}{\epsilon}\Big(2(\lambda_1 + \lambda_2)\lambda_4 + 2\lambda^2_4 + 4\lambda_3\lambda_4 + 4|\lambda_5|^2 + \frac{3}{2} g^2{g'}^2 \nonumber \\ 
& + 5( |\lambda_6|^2+ |\lambda_7|^2) + 2 \RE(\lambda_6 \lambda_7) \Big), \quad  \\
\delta \lambda_{5}\,=& \frac{1}{16\pi^2}\frac{1}{\epsilon}\Big( 2(\lambda_1 + \lambda_2 + 2 \lambda_3 + 3 \lambda_4)\lambda_5 + 5(\lambda_6 \lambda_6 + \lambda^*_7 \lambda^*_7) + 2 \lambda_6 \lambda^*_7 \Big), \\
\delta \lambda_{6}\,=& \frac{1}{16\pi^2}\frac{1}{\epsilon}\Big( 12 \lambda_1 \lambda_6 + (3\lambda_3 + 2\lambda_4) \lambda^*_7 + \lambda_5\lambda_7 + (3\lambda_3 + 4\lambda_4)\lambda_6 + 5 \lambda_5 \lambda^*_6  \Big), \\
\delta \lambda_{7}\,=& \frac{1}{16\pi^2}\frac{1}{\epsilon}\Big( 12 \lambda_2 \lambda_7 + (3\lambda_3 + 2\lambda_4) \lambda^*_6 + \lambda^*_5\lambda_6 + (3\lambda_3 + 4\lambda_4)\lambda_7 + 5 \lambda^*_5 \lambda^*_7  \Big).
\end{align}
These may also be used to renormalize the theory in vacuum and are thus useful for determining one-loop-corrected relations to physical pole masses. We leave this calculation for future work.

By requiring that the bare parameters are independent of the renormalization scale, one obtains the following $\beta$ functions:
\begin{align}
\label{running_g}
\Lambda \frac{d}{d\Lambda}g^2 ={}& - \frac{g^4}{8 \pi^2} \bigg( \frac{22}{3}-\frac{N_d}{6}-\frac{4}{3}N_f \bigg), \\
\Lambda \frac{d}{d\Lambda}g'^2 ={}& \frac{g'^4}{8 \pi^2} \bigg( \frac{N_d}{6} + \frac{20}{9}N_f \bigg), \\
\Lambda \frac{d}{d\Lambda}g^2_{Y} ={}& \frac{g^2_{Y}}{8 \pi^2} \bigg( \frac{9}{2}g^2_{Y} - \frac{9}{4}g^2 - \frac{17}{12}g'^2 - 8 g^2_s \bigg), \\
\Lambda \frac{d}{d\Lambda}\mu^2_{11} ={}& \frac{1}{16\pi^2} \bigg(-3\mu^2_{11}\Big(\frac{3}{2}g^2 +\frac{1}{2}{g'}^2 - 4 \lambda_1 \Big) + 2 \mu^2_{22}(2\lambda_3 + \lambda_4) + 12\RE(\lambda^*_6 \mu^{2}_{12}) \bigg), \\
\Lambda \frac{d}{d\Lambda}\mu^2_{22} ={}& \frac{1}{16\pi^2} \bigg(-3\mu^2_{22}\Big(\frac{3}{2}g^2 +\frac{1}{2}{g'}^2 - 2 g^2_{Y} - 4 \lambda_2 \Big) + 2 \mu^2_{11}(2\lambda_3 + \lambda_4) + 12\RE(\lambda_7 \mu^{2}_{12}) \bigg), \\
\Lambda \frac{d}{d\Lambda}\mu^2_{12} ={}& \frac{1}{16\pi^2} \bigg(-3\mu^2_{12}\Big(\frac{3}{2}g^2 +\frac{1}{2}{g'}^2 - g^2_{Y}  \Big) + 6(\mu^2_{11}\lambda_6 + \mu^2_{22}\lambda^*_7) + 6\lambda_5 \mu^{2*}_{12} \nonumber \\
& + 2(2\lambda_4+\lambda_3)\mu^2_{12} \bigg), \\
\Lambda \frac{d}{d\Lambda}\lambda_1 ={}& \frac{1}{16\pi^2} \frac{1}{2} \bigg(48\lambda^2_1 + 4 \lambda^2_3 + 4\lambda_3 \lambda_4 + 2\lambda^2_4 + 2|\lambda_5|^2 + 24|\lambda_6|^2 +\frac{3}{4}(3g^4 + {g'}^4 + 2 g^2{g'}^2)  \nonumber \\ 
& -6\lambda_1 (3g^2 + {g'}^2 ) \bigg), \\
\Lambda \frac{d}{d\Lambda}\lambda_2 ={}& \frac{1}{16\pi^2}\frac{1}{2} \bigg(48\lambda^2_2 + 4 \lambda^2_3 + 4\lambda_3 \lambda_4 + 2\lambda^2_4 + 2|\lambda_5|^2 + 24|\lambda_7|^2 +\frac{3}{4}(3g^4 + {g'}^4 + 2 g^2{g'}^2) \nonumber \\
& - 12 g^4_{Y} -6\lambda_2 (3g^2 + {g'}^2 - 4 g^2_{Y}) \bigg), \\
\Lambda \frac{d}{d\Lambda}\lambda_3 ={}& \frac{1}{16\pi^2} 2\bigg(2(\lambda_1 + \lambda_2)(3\lambda_3 + \lambda_4) + 2\lambda^2_3 +  \lambda^2_4 + |\lambda_5|^2 + \frac{3}{8}(3g^4 + {g'}^4 - 2 g^2{g'}^2) \nonumber \\ & + 2( |\lambda_6|^2+ |\lambda_7|^2) + 8 \RE(\lambda_6 \lambda_7) - \frac{3}{2}\lambda_3 \Big(3g^2+{g'}^2 - 2g^2_{Y}\Big) \bigg), \\
\Lambda \frac{d}{d\Lambda}\lambda_4 ={}& \frac{1}{16\pi^2} 2 \bigg(2(\lambda_1 + \lambda_2)\lambda_4 + 2\lambda^2_4 + 4\lambda_3\lambda_4 + 4|\lambda_5|^2 + \frac{3}{2} g^2{g'}^2 \nonumber \\ 
& + 5( |\lambda_6|^2+ |\lambda_7|^2) + 2 \RE(\lambda_6 \lambda_7) - \frac{3}{2}\lambda_4 \Big(3g^2+{g'}^2 - 2g^2_{Y}\Big)  \bigg), \\
\Lambda \frac{d}{d\Lambda}\lambda_5 ={}& \frac{1}{16\pi^2} 2 \bigg( 2(\lambda_1 + \lambda_2 + 2 \lambda_3 + 3 \lambda_4)\lambda_5 + 5(\lambda_6 \lambda_6 + \lambda^*_7 \lambda^*_7) + 2 \lambda_6 \lambda^*_7  \nonumber \\ & - \frac{3}{2}\lambda_5 \Big(3g^2+{g'}^2 - 2g^2_{Y}\Big) \bigg), \\
\Lambda \frac{d}{d\Lambda}\lambda_6 ={}& \frac{1}{16\pi^2} 2 \bigg( 12 \lambda_1 \lambda_6 + (3\lambda_3 + 2\lambda_4) \lambda^*_7 + \lambda_5\lambda_7 + (3\lambda_3 + 4\lambda_4)\lambda_6 + 5 \lambda_5 \lambda^*_6  \nonumber \\ 
& - \frac{3}{2}\lambda_6 \Big(3g^2+{g'}^2 - g^2_{Y}\Big) \bigg), \\
\Lambda \frac{d}{d\Lambda}\lambda_7 ={}& \frac{1}{16\pi^2} 2 \bigg(12 \lambda_2 \lambda_7 + (3\lambda_3 + 2\lambda_4) \lambda^*_6 + \lambda^*_5\lambda_6 + (3\lambda_3 + 4\lambda_4)\lambda_7 + 5 \lambda^*_5 \lambda^*_7  \nonumber \\ 
& - \frac{3}{2}\lambda_7 \Big(3g^2+{g'}^2 - 3g^2_{Y}\Big) \bigg).
\end{align}
Two-loop-corrected $\beta$ functions have been obtained in Ref.~\cite{Dev:2014yca}.

\subsection{One-loop thermal masses}
\label{sec:thermal_masses}

Here we collect the one-loop thermal masses that are needed for thermal counterterms in the four-dimensional theory:
\begin{align}
\bar{\Pi}_{1}  \equiv& \frac{T^2}{12}\Big(6\lambda_1 + 2 \lambda_3 + \lambda_4 + \frac{d}{4}(3g^2 + {g'}^2)  \Big), \\
\bar{\Pi}_{2} \equiv& \frac{T^2}{12}\Big(6\lambda_2 + 2 \lambda_3 + \lambda_4 + \frac{d}{4}(3g^2 + {g'}^2) -6(2^{2-d}-1) g^2_{Y} \Big), \\
\bar{\Pi}_{12} \equiv& \frac{3 T^2}{12} (\lambda_6+\lambda_7^*), \\
m_D^2=&{}g^2T^2\bigg(\frac{4+N_d}{6}+\frac{N_f}{3}\bigg),\\
m'^2_D=&{}g'^2T^2\bigg(\frac{N_d}{6}+\frac{5N_f}{9}\bigg). 
\end{align} 

In the effective theory containing temporal scalar fields $A_0, B_0$ and $C_0$, the analogous mass corrections read
\begin{align}
\bar{\Pi}_{\phi_1,3} &\equiv - \frac{m_D}{4\pi} (3h_{1} + h_{2}), \\
\bar{\Pi}_{\phi_2,3} &\equiv - \frac{m_D}{4\pi} (3h_{4} + h_{5}).
\end{align}
Contributions from temporal gluons are of higher order, and have been omitted.

After the temporal scalars have been integrated out, the mass correction for the $\phi$ field in the diagonalized theory is given by
\begin{align}
\bar{\Pi}_{\phi,3} = -\frac{m_\theta}{4\pi}(2 \tilde{\lambda}_3+\tilde{\lambda}_4).
\end{align}

\subsection{Normalization of fields}

Relations between four- and three-dimensional fields, in Landau gauge, read:

\begin{align}
A_{\td,0}^2 &=\frac{A_{\fd,0}^2}{T}\bigg[1+\frac{g^2}{(4\pi)^2}\Big(\frac{N_d-26}{6}L_b+\frac{1}{3}(8+N_d)+\frac{4N_f}{3}(L_f-1)\Big)\bigg],\\
A_{\td,r}^2 &=\frac{A_{\fd,r}^2}{T}\bigg[1+\frac{g^2}{(4\pi)^2}\bigg(\frac{N_d-26}{6}L_b-\frac{2}{3}+\frac{4N_f}{3}L_f\bigg)\bigg], \\
B_{\td,0}^2
&=\frac{B_{\fd,0}^2}{T}\bigg[1+\frac{g'^2}{(4\pi)^2}\Big(N_d\Big(\frac{L_b}{6}+\frac{1}{3}\Big)+\frac{20N_f}{9}(L_f-1)\Big)
\Big],\\ B_{\td,r}^2 &=\frac{B_{\fd,r}^2}{T}\bigg[1+
  \frac{g'^2}{(4\pi)^2}\bigg(N_d\frac{L_b}{6}+\frac{20N_f}{9}L_f\bigg)\bigg], \\
\big(\phi^{\dagger}_1\phi_1\big)_\td &=\frac{\big(\phi^{\dagger}_1\phi_1\big)_\fd}{T}\bigg[1-\frac{1}{(4\pi)^2}\Big(\frac{3}{4}(3g^2 + {g'}^2)L_b  \Big)\bigg], \\
\big(\phi^{\dagger}_2\phi_2\big)_\td &=\frac{\big(\phi^{\dagger}_2\phi_2\big)_\fd}{T}\bigg[1-\frac{1}{(4\pi)^2}\Big(\frac{3}{4}(3g^2 + {g'}^2)L_b - 3 g^2_{Y} L_f  \Big)\bigg], \\
\big(\phi^{\dagger}_1\phi_2\big)_\td &=\frac{\big(\phi^{\dagger}_1\phi_2\big)_\fd}{T}\bigg[1-\frac{1}{(4\pi)^2}\Big(\frac{3}{4}(3g^2 + {g'}^2)L_b - \frac{3}{2} g^2_{Y} L_f \Big)\bigg].
\end{align}

\subsection{Mass counterterms in the effective theories}
\label{sec:mass_ct}

We list here mass counterterms in the effective theories. These play a role in determining relations between lattice and continuum physics \cite{Laine:1995np,Laine:1997dy}. In the first effective theory containing temporal scalars (Eq.~(\ref{eq:3d_heavy})), UV divergences are canceled by introducing counterterms as
\begin{align}
\delta\mu^2_{22,3} =& -\frac{1}{16\pi^2}\frac{1}{4\epsilon}\Big( \frac{39 g_3^4}{16} -\frac{5}{16} g_3'{}^4 -\frac{9}{8} g_3^2 g_3'{}^2 +3 (3g_3^2 + g_3'{}^2) \lambda _{2,3}  -12 \lambda
   _{2,3}^2 +12 g_3^2 h_4 \nonumber \\ 
&-6 h_4^2-2 h_5^2-3 h_6^2\Big)_{\text{SM}} 
 -\frac{1}{16\pi^2}\frac{1}{4\epsilon} \Big( -\frac{1}{8} \big(3g_3^4 + g_3'{}^4\big) + (3g_3^2  +g_3'{}^2) (\lambda_{3,3} + \frac12 \lambda_{4,3})  \nonumber \\
& -2 ( \lambda _{3,3}^2 + \lambda _{3,3} \lambda _{4,3} + \lambda _{4,3}^2) - 3 |\lambda_{5,3}|^2 - 3 |\lambda_{6,3}|^2 - 9 |\lambda_{7,3}|^2 \Big)_\text{2HDM} \, , \\
\delta\mu^2_{11,3} =& -\frac{1}{16\pi^2}\frac{1}{4\epsilon}\Big( \frac{33 g_3^4}{16}-\frac{7}{16} g_3'{}^4 -\frac{9}{8} g_3^2 g_3'{}^2  +3 (3g_3^2 + g_3'{}^2) \lambda _{1,3}  \nonumber \\ 
& +( 3g_3^2 +g_3'{}^2) (\lambda _{3,3} + \frac12 \lambda _{4,3}) -12 \lambda _{1,3}^2 -2 (\lambda _{3,3}^2 + \lambda _{3,3} \lambda _{4,3}+ \lambda_{4,3}^2) \nonumber \\ 
&- 3 |\lambda_{5,3}|^2 - 9 |\lambda_{6,3}|^2 - 3 |\lambda_{7,3}|^2 +12 g_3^2 h_1-6h_1^2-2 h_2^2-3 h_3^2 \Big), \\
\delta\mu^2_{12,3} =& -\frac{1}{16\pi^2}\frac{1}{4\epsilon}\Big( \frac32 \big( 3 g_3^2 + g'{}^2\big) \big(\lambda_{6,3} + \lambda_{7,3}^* \big) - 3(2 \lambda_{1,3}+\lambda_{3,3}+\lambda_{4,3}) \lambda_{6,3}  \nonumber \\ 
&  - 3(2 \lambda_{2,3}+\lambda_{3,3}+\lambda_{4,3}) \lambda_{7,3}^* - 3 \lambda_{5,3}(\lambda_{6,3}^* + \lambda_{7,3}) \Big).
\end{align}
For convenience, we have separated contributions from diagrams specific to 2HDM in the equation for $\delta\mu^2_{22,3}$.

In the effective theory of Eq.~(\ref{eq:lag_3d_scalars}) where temporal scalars have been integrated out, the mass counterterms are given by 
\begin{align}
\delta\bar{\mu}^2_{22,3} =& \delta\mu^2_{22,3}\Big\vert_{h_i = 0} -\frac{1}{16\pi^2}\frac{1}{4\epsilon}\Big( \frac{3}{16} \bar{g}^4_3 \Big)_{\text{SM}} , \\
\delta\bar{\mu}^2_{11,3} =& \delta\mu^2_{11,3}\Big\vert_{h_i = 0} -\frac{1}{16\pi^2}\frac{1}{4\epsilon}\Big( \frac{3}{16} \bar{g}^4_3 \Big), \\
\delta\bar{\mu}^2_{12,3} =& \delta\mu^2_{12,3}, 
\end{align}
where the parameters are understood to be $\bar{g_3}, \bar{g}'_3, \bar{\lambda}_{1,3}\dots$. 

Finally, the mass counterterm in the SM-like effective theory reads 
\begin{align}
\delta\hat{\mu}^2_{3} =& -\frac{1}{16\pi^2}\frac{1}{4\epsilon}\Big( \frac{51 \hat{g}_3^4}{16} -\frac{5}{16} \hat{g}_3'{}^4 -\frac{9}{8} \hat{g}_3^2 \hat{g}_3'{}^2 +3 (3\hat{g}_3^2 + \hat{g}_3'{}^2) \hat{\lambda}_{3}  -12 \hat{\lambda}_{3}^2\Big).
\end{align}

\subsection{Collection of integrals}
\label{sec:integrals}

The Euclidean four-momentum is denoted as $P=(\omega_n,\vek p)$ for bosons, where $\omega_n=2n\pi T$, and as $P=(\nu_n,\vek p)$ for fermions, where $\nu_n=(2n+1)\pi T$. 
In dimensional regularization, spatial integration is performed in $d\equiv3-2\eps$ dimensions. We introduce the following shorthand notation for the combined Matsubara sum and space integration:
\begin{equation}
\begin{split}
\text{bosons:}&\quad\sumint P\equiv T\sum_{\omega_n}\int_p,\\
&\quad\sumint P'\equiv T\sum_{\omega_n\neq0}\int_p\qquad\qquad\text{(sum over nonzero modes)},\\
\text{fermions:}&\quad\sumint{\braced P}\equiv T\sum_{\nu_n}\int_p,\quad
\text{where}\quad\int_p\equiv\xifac\int\frac{\dd^{3-2\eps}\vek p}{(2\pi)^{3-2\eps}}.
\end{split}
\end{equation}
All integrals relevant for $O(g^4)$ DR are listed below. 

\subsubsection{Three-dimensional integrals}

We denote
\begin{align}
\mathcal P_T(k)_{rs}&\equiv\delta_{rs}-\frac{k_r k_s}{k^2}.
\end{align}

\paragraph{One-loop integrals}

\begin{align}
I^3_\alpha(m) &\equiv \int_p \frac{1}{(p^2+m^2)^\alpha} =  \bigg(\frac{e^\gamma \Lambda^2}{4\pi}\bigg)^\epsilon \frac{(m^2)^{\frac{d}{2}-\alpha}}{(4\pi)^\frac{d}{2}}\frac{\Gamma(\alpha-\frac{d}{2})}{\Gamma(\alpha)} , \\
L^3_2(m_1,m_2) &\equiv \int_p \frac{1}{(p^2+m^2_1)(p^2+m^2_2)} = \frac{1}{m^2_2-m^2_1}\Big( I^3_1(m_1)-I^3_1(m_2) \Big), \\
I^3_1(m) &=\int_p\frac1{p^2+m^2}={}-\frac m{4\pi}\mutwom2\left[1+2\eps+\OO(\eps^2)\right],\\
I^3_2(m) &=\int_p\frac1{(p^2+m^2)^2}={}\frac1{8\pi m}\mutwom2\left[1+\OO(\eps^2)\right].
\end{align}

\paragraph{Two-loop integrals}
\begin{align}
& \int_{pq}\frac1{(p^2+m^2)^\alpha(q^2+m^2)^\beta[(\vek p+\vek q)^2]^\delta} \nonumber \\
& =\Big(\frac{e^{\gamma}\Lambda^2}{4\pi}\Big)^{2\epsilon}\frac{(m^2)^{d-\alpha-\beta-\delta}}{(4\pi)^d} 
\frac{\Gamma\left(\frac d2-\delta\right)\Gamma\left(\alpha+\delta-\frac d2\right)\Gamma\left(\beta+\delta-\frac d2\right)\Gamma(\alpha+\beta+\delta-d)}{\Gamma\left(\frac d2\right)\Gamma(\alpha)\Gamma(\beta)\Gamma(\alpha+\beta+2\delta-d)}, \\
&\int_{pq}\frac1{(p^2+m^2)(q^2)^\alpha[(\vek p-\vek q)^2]^\beta} \nonumber \\
& = \Big(\frac{e^{\gamma}\Lambda^2}{4\pi}\Big)^{2\epsilon}\frac{(m^2)^{d-\alpha-\beta-1}}{(4\pi)^d}
\frac{\Gamma\left(1+\alpha+\beta-d\right) \Gamma\left(\alpha+\beta-\frac{d}{2} \right) \Gamma\left(\frac{d}{2}-\alpha \right) \Gamma\left(\frac{d}{2}-\beta \right)}{\Gamma\left(\alpha \right)\Gamma\left(\beta \right)\Gamma\left(\frac{d}{2} \right)}, \\
\label{eq:sunset_equal}
&\int_{pq} \frac{1}{(p^2+m^2)(q^2+m^2)[(\vek p+\vek q)^2+m^2]} \nonumber \\
& = -\Big(\frac{1}{2\pi} \Big)^{2d} \Big(\frac{e^\gamma \Lambda^2}{4\pi} \Big)^{2\epsilon} \frac{3(d-2)}{4(d-3)} (m^2)^{d-3} \Big(\pi^{\frac{d}{2}} \Gamma(1-\frac{d}{2}) \Big)^2 \nonumber \\
& \quad \times \bigg( {}_2 F_1\Big(\frac{4-d}{2},1;\frac{5-d}{2};\frac{3}{4} \Big) - 3^{\frac{d-5}{2}} 2\pi \frac{\Gamma(5-d)}{\Gamma(\frac{4-d}{2})\Gamma(\frac{6-d}{2})} \bigg).
\end{align}
Eq.~(\ref{eq:sunset_equal}) can be found in Ref.~\cite{Schroder:2005va}. For arbitrary masses, the three-dimensional sunset integral has a series expansion given by
\begin{align}
S^3_3(m_1,m_2,m_3) &\equiv \int_{p,q} \frac{1}{(p^2+m^2_1)(k^2+m^2_2)((\vek p + \vek k)^2+m^2_3)} \nonumber \\
&= \frac{1}{16\pi^2}\bigg(\frac{1}{4\epsilon} + \ln\Big(\frac{\Lambda}{m_1+m_2+m_3}\Big) + \frac{1}{2} \bigg),
\end{align}
and we need the following integrals containing gauge-field propagators: 
\begin{align}
B^3_4(m) \equiv& \int_{pq} \frac{(2p+q)_r(2p+q)_s}{(p^2+m^2)^2 q^2 ((p+q)^2+m^2)} \mathcal P_T(q)_{rs} = \frac{1}{16\pi^2}\frac{1}{2} \bigg(\frac{1}{\epsilon} + 1 + 4 \ln\Big(\frac{\Lambda}{2m} \Big) \bigg), \\
B^3_5(m) \equiv& \int_{pq}  \frac{(p-q)_r(p-q)_s}{(p^2+m^2)(q^2+m^2)(p+q)^4} \mathcal P_T(p+q)_{rs}  = -\frac{1}{16\pi^2} \frac{1}{4}\bigg(\frac{1}{\epsilon} + 4\ln\Big(\frac{\Lambda}{2m} \Big) \bigg),\\
B^3_6(m) \equiv& \int_{pq} \frac{ \mathcal P_\tau(p)_{rs} \mathcal P_T(q)_{rs}}{p^2 q^2 ((p+q)^2+m^2)} = \frac{1}{16\pi^2}\bigg(\frac{3}{8\epsilon} +\frac{1}{8}+ \frac{3}{2}\ln\Big(\frac{\Lambda}{m} \Big) \bigg).
\end{align}

Finally, we have extracted the UV divergent parts of the following integrals that vanish in dimensional regularization due to exact cancellation of UV and IR divergences:
\begin{align}
b^3_3(m) &\equiv \int_{pq}\frac{q_r (p_s+q_s) \mathcal P_\tau(p)_{rs} }{(p^2+m^2)^2q^2)[(\vek p+\vek q)^2]^2}={} -\frac{1}{16\pi^2} \frac{1}{16\epsilon} + \text{UV finite part}, \\
b^3_7(m) &\equiv \int_{pq} \frac{ \mathcal P_\tau(p)_{rs} \mathcal P_\tau(q)_{ij} \mathcal P_\tau(p+q)_{mn} (q_s \delta_{mi} - p_i \delta_{ms} - q_m \delta_{is} ) (q_r \delta_{nj} - p_j \delta_{rn} - q_n \delta_{rj} )}{(p^2+m^2)^2(q^2+m^2)[(\vek p+\vek q)^2+m^2]} \nonumber \\
&={} \frac{1}{16\pi^2} \frac{5}{16\epsilon} + \text{UV finite part}.
\end{align} 
These are needed in the limit $m \rightarrow 0$ to calculate mass counterterms in the effective theories.

\subsubsection{Four-dimensional sum-integrals}

\begin{align}
I^{4b}_{\alpha,\beta,\delta} &\equiv\,\sumint P'\frac{(P_0^2)^\beta(\vek p^2)^\delta}{(P^2)^\alpha}={} \nonumber \\
& \frac{(e^\gamma\Lambda^2)^\eps}{8\pi^2}\frac{\Gamma\left(\alpha-\tfrac d2-\delta\right)\Gamma\left(\tfrac d2+\delta\right)\zeta(2\alpha-2\beta-2\delta-d)}{\Gamma\left(\tfrac12\right)\Gamma(\alpha)\Gamma\left(\tfrac d2\right)}(2\pi T)^{1+d-2\alpha+2\beta+2\delta},\\
I^{4b}_{\alpha,\beta} &\equiv I^{4b}_{\alpha,\beta,0}=\,\sumint P'\frac{(P_0^2)^\beta}{(P^2)^\alpha}={}\frac{(e^\gamma\Lambda^2)^\eps}{8\pi^2}\frac{\Gamma\left(\alpha-\tfrac d2\right)\zeta(2\alpha-2\beta-d)}{\Gamma\left(\tfrac12\right)\Gamma(\alpha)}(2\pi T)^{1+d-2\alpha+2\beta},\\
I^{4b}_\alpha &\equiv I^{4b}_{\alpha,0}=\,\sumint{P}'\frac{1}{(P^2)^\alpha},\,\phantom{=}&\\
I^{4b}_1 &=\,\sumint P'\frac1{\phantom(P^2\phantom{)^\alpha}}={}\frac{T^2}{12}\mufourpiT2\left\{1+2\left[\log2\pi+\gamma-\frac{\zeta'(2)}{\zeta(2)}\right]\eps+\OO(\eps^2)\right\},\\
I^{4b}_2 &=\,\sumint P'\frac1{(P^2)^2}={}\frac1{16\pi^2}\mufourpiT2\left[\frac1\eps+2\gamma+\OO(\eps)\right].
\end{align}

For the one-loop effective potential we need the sum-integrals
\begin{align}
\label{eq:Jb}
J_b(m) &= \frac{1}{2} \sumint K \log(K^2+m^2) \nonumber \\
&= \frac{m^2T^2}{24} - \frac{m^3T}{12\pi}-\frac{m^4}{64\pi^2}\left(\frac{1}{\epsilon}+L_b
\right) + \frac{\zeta(3)m^6}{3(4\pi)^4T^2} + \mathcal{O}\left(\frac{m^8}{T^2},\epsilon\right), \\
\label{eq:Jf}
J_f(m) &= \frac{1}{2} \sumint {\{K\}} \log(K^2+m^2) \nonumber \\
&= -\frac{m^2T^2}{48} -\frac{m^4}{64\pi^2}\left(\frac{1}{\epsilon}+L_f
\right) + \frac{7\zeta(3)m^6}{3(4\pi)^4T^2} + \mathcal{O}\left(\frac{m^8}{T^2},\epsilon\right).
\end{align}

\paragraph{Two-loop sum-integrals}\mbox{}\\
We introduce the following shorthand notation:
\begin{align}
\underline{m} &\equiv \sqrt{m^2 + m^2_T }, \\
\mathcal{S}(P,\underline{m}) &\equiv \frac{1}{P^2+m^2+\delta_{P_0}m^2_T}, \\
\mathcal P_T(K)_{\mu\nu}&\equiv\delta_{\mu\nu}-\frac{K_\mu K_\nu}{K^2},  \\
\mathcal P_\tau(K)_{\mu\nu}&\equiv \delta_{\mu i }\delta_{\nu j} \Big(\delta_{ij}-\frac{k_i k_j}{\vek k^2}\Big), \\
D^{\mu\nu}_\alpha(K,m_T) &\equiv \bigg( (1-\delta_{K_0})\frac{\mathcal P_T(K)_{\mu\nu}}{(K^2)^\alpha} + \delta_{K_0}\Big(\frac{\delta_{\mu 0} \delta_{\nu 0}}{(K^2+m^2_T)^\alpha} + \frac{\mathcal P_\tau(K)_{\mu\nu}}{(K^2)^\alpha} \Big)  \bigg).
\end{align}
$m^2_T$ corresponds to a thermal-mass correction required for resummation. 

Results for the sum-integrals are given up to terms of order $O(|m|T)\sim O(gT^2)$ in high-$T$ expansion.
\begin{align}
F_{1}(m) \equiv& \sumint{\braced P,K} \frac{D^{\mu\mu}_1(K,m)}{P^2(P+K)^2}  \simeq 0, \\ \nonumber \\
F_{2}(m) \equiv& \sumint{\braced P,K} \frac{D^{\mu\nu}_1(K,m)}{P^4(P+K)^2}\Big(-\delta_{\mu\nu}(P^2 + P \cdot K) + 2 P_\mu P_\nu \Big) \nonumber \\
& \simeq \frac{1}{2}(1-d)(2-2^{2-d})(2^{4-d}-1) I^{4b}_1 I^{4b}_2 , \\ \nonumber \\
F_{3}(m) \equiv& \sumint{\braced P,K} \frac{D^{\mu\nu}_2(K,m)}{P^2(P+K)^2}\Big(-\delta_{\mu\nu}(P^2 + P \cdot K) + 2 P_\mu P_\nu  \Big) \nonumber \\
& \simeq (1-d)(2^{2-d}-1)\Big( T I^{4b}_1 I^3_2(m) + I^{4b}_1 I^{4b}_2 \Big) , \\ \nonumber \\
F_{4}(\underline{m}) \equiv& \sumint{\braced P,K} \frac{P^2 + P \cdot K }{P^4 (P+K)^2} \mathcal{S}(K,\underline{m}) \nonumber \\
& \simeq \frac{1}{2}(2-2^{2-d})(2^{4-d}-1) I^{4b}_1 I^{4b}_2  ,\\ \nonumber \\
F_{5}(\underline{m}) \equiv& \sumint{\braced P,K} \frac{P^2 + P \cdot K}{P^2 (P+K)^2} \Big( \mathcal{S}(K,\underline{m}) \Big)^2 \nonumber \\
& \simeq (2^{2-d}-1)\Big(  I^{4b}_1 I^{4b}_2 + T  I^{4b}_1 I^3_2(\underline{m})  \Big) , \\ \nonumber \\
S_{1}(\underline{m}_1,\underline{m}_2) \equiv& \sumint{P,K} \Big(\mathcal{S}(P,\underline{m}_1)\Big)^2 \mathcal{S}(K,\underline{m}_2)  \nonumber \\ 
& \simeq T^2 I^3_2(\underline{m}_1) I^3_1(\underline{m}_2)  +T I^3_2(\underline{m}_1) I^{4b}_1 + I^{4b}_1 I^{4b}_2, \\ \nonumber \\
S_{2}(\underline{m}_1,\underline{m}_2) \equiv& \sumint{P,K} \mathcal{S}(P,\underline{m}_1) \mathcal{S}(P,\underline{m}_2) \mathcal{S}(K,\underline{m}_2) \nonumber \\ 
& \simeq T^2 L^3_2(\underline{m}_1,\underline{m}_2) I^3_2(\underline{m}_2) + T L^3_2(\underline{m}_1,\underline{m}_2) I^{4b}_1 + I^{4b}_1 I^{4b}_2, \\ \nonumber \\
S_3(\underline{m}_1,\underline{m}_2,\underline{m}_3) \equiv& \sumint{P,K}  \mathcal{S}(P,\underline{m}_1)  \mathcal{S}(K,\underline{m}_2)  \mathcal{S}(P+K,\underline{m}_3) \nonumber \\ 
&\simeq T^2 S^3_3(\underline{m}_1,\underline{m}_2,\underline{m}_3), \\ \nonumber \\
B_{11}(\underline{m}_1,m_2) \equiv& \sumint{P,K} D^{\mu\mu}_1(K,m_2) \Big(\mathcal{S}(P,\underline{m}_1)\Big)^2 \nonumber \\ 
&\simeq  T^2 I^3_1(m_2) I^3_2(\underline{m}_1) + d \Big(T I^3_2(\underline{m}_1) I^{4b}_1 + I^{4b}_1 I^{4b}_2\Big) , \\ \nonumber \\ 
B_{12}(\underline{m}_1,m_2) \equiv& \sumint{P,K}  D^{\mu\mu}_2(K,m_2) \mathcal{S}(P,\underline{m}_1) \nonumber \\ 
&\simeq   T^2 I^3_1(\underline{m}_1) I^3_2(m_2) +  T I^3_2(m_2) I^{4b}_1 + d I^{4b}_1 I^{4b}_2, \\ \nonumber \\ 
L_0 \equiv &  \sumint{P,K} \frac{(P \cdot K)^2}{P^4 K^6}
= \frac{1}{8}(d-2)(d-4)I^{4b}_1 I^{4b}_2, \\ \nonumber \\
B_2(m) \equiv& \sumint{P,K} (-2 \delta_{\mu\nu}\delta_{\rho\sigma} + \delta_{\mu\sigma}\delta_{\nu\rho} + \delta_{\mu\rho}\delta_{\nu\sigma}) D^{\mu\nu}_2(P,m) D^{\rho\sigma}_1(K,m) \nonumber \\
&\simeq 2 L_0  - d  T I^3_2(m) I^{4b}_1 + 2(d -d^2 -1)I^{4b}_1 I^{4b}_2,\\ \nonumber \\ 
B_3(m) \equiv& \sumint{P,K} \frac{K_\mu K_\nu}{K^2(P+K)^2} D^{\mu\nu}_2(P,m) \nonumber \\
&\simeq \left(1-\frac{d}{2}\right) T I^3_2(m) I^{4b}_1 + \frac{1}{2} I^{4b}_1 I^{4b}_2,  \\ \nonumber \\ 
B_4(\underline{m}_1,m_2) \equiv& \sumint{P,K} (2P+K)_\mu(2P+K)_\nu  D^{\mu\nu}_1(K,m_2) \Big( \mathcal{S}(P,\underline{m}_1) \Big)^2 \nonumber \\ 
& \quad \quad \times \mathcal{S}(P+K,m_1,m_{T}) \nonumber \\  
&\simeq T^2 B^3_4(\underline{m}_1),  \\ \nonumber \\ 
B_5(\underline{m}_1,m_2) \equiv& \sumint{P,K} (P-K)_\mu(P-K)_\nu  D^{\mu\nu}_2(P+K,m_2)  \mathcal{S}(P,\underline{m}_1)  \mathcal{S}(K,\underline{m}_1) \nonumber \\ 
&\simeq T^2 B^3_5(\underline{m}_1) +4\left(1-\frac{d}{2}\right) T I^3_2(m_2) I^{4b}_1+ 2 I^{4b}_1 I^{4b}_2 \\ \nonumber \\ 
B_6(\underline{m}_1,m_2,m_3) \equiv& \sumint{P,K}  D^{\mu\nu}_1(P,m_2) D^{\mu\nu}_1(K,m_3) \mathcal{S}(P+K,\underline{m}_1) \nonumber \\ 
&\simeq T^2 \Big( B^3_6(\underline{m}_1) + S^3_3(\underline{m}_1,m_2,m_3) \Big)  ,\\ \nonumber \\  
B_7(m) \equiv& \sumint{P,K}  D^{\beta\nu}_2(P,m)  D^{\kappa\lambda}_1(K,m)  D^{\rho\sigma}_1(P+K,m) \nonumber \\
& \quad \quad \quad \times (K_\nu \delta_{\rho\kappa} - P_\kappa \delta_{\rho\nu} - K_\rho \delta_{\kappa\nu} ) (K_\beta \delta_{\sigma\lambda} - P_\lambda \delta_{\beta\sigma} - K_\sigma \delta_{\beta\lambda} ) \nonumber \\
&\simeq \frac{1}{4} T^2 \Big(2B^3_4(m,m) + B^3_5(m,m) \Big) + \frac{1}{4}(1+2d) I^{4b}_1 I^{4b}_2 \nonumber \\
& \quad \quad - \frac{1}{2} L_0  +\frac{d}{2}(d-2) \; T I^3_2(m) I^{4b}_1.
\end{align}
Finally, note that the following fermionic sunset sum-integral vanishes:
\begin{align}
& \sumint{\braced P,K} \frac{1}{P^2K^2(P+K)^2} = 0.
\end{align}

These master sum-integrals are required in the evaluation of two-loop scalar two-point functions directly in the unbroken phase, without using the effective potential (which was used in Ref.~\cite{Kajantie:1995dw}).  It will turn out that terms with mixed zero mode and non-zero mode contributions, i.e. terms of the form $I^3_2(m) I^{4b}_1$, are entirely canceled in resummation, and so the matching relations are obtained solely from the pure non-zero mode parts.

\subsection{Effective potential for the SM-like theory with a 6-dim. operator }
\label{sec:dim6-Veff}

We describe here details of the calculation leading to the error estimate discussed in Section \ref{sec:dim6-step3}. The 6-dim. operator $\hat{\Lambda}_6 (\phi^\dagger \phi)$ modifies couplings in the mass-eigenstate basis, and also enters the relations for mass eigenvalues. Parameterizing the doublet $\phi$ in the effective theory as 
\begin{equation}
\phi= 
\begin{pmatrix}
	G^+ \\
	\frac{1}{\sqrt{2}} (\varphi + h + i G)
\end{pmatrix},
\end{equation}
where $\varphi$ is a classical background field, the scalar masses read
\begin{align}
m_h^2&=3 \hat{\lambda }_3 \varphi ^2+\frac{15}{4} \hat{\Lambda }_6 \varphi ^4+\hat{\mu }_3^2, \\
m_G^2= m_{G^{\pm}}^2 &=\hat{\lambda }_3 \varphi ^2+\frac{3}{4} \hat{\Lambda }_6 \varphi ^4+\hat{\mu }_3^2,
\end{align}
while the gauge boson masses obtain no corrections from $\hat{\Lambda}_6$. Furthermore, the 6-dim. operator contributes to vertex rules of the scalar sector as follows:
\begin{align}
C_{h,h,h}&=-3 \left(2 \hat{\lambda }_3 \varphi +5 \hat{\Lambda }_6 \varphi ^3\right), \\
C_{h,h,h,h}&=-6 \hat{\lambda }_3-45 \hat{\Lambda }_6 \varphi ^2, \\
C_{h,h,G,G}&=-2 \hat{\lambda }_3-9 \hat{\Lambda }_6 \varphi ^2, \\
C_{h,h,G^+,G^-}&=-2 \hat{\lambda }_3-9 \hat{\Lambda }_6 \varphi ^2, \\
C_{h,G,G}&=-2 \hat{\lambda }_3 \varphi -3 \hat{\Lambda }_6 \varphi ^3, \\
C_{h,G^+,G^-}&=-2 \hat{\lambda }_3 \varphi -3 \hat{\Lambda }_6 \varphi ^3, \\
C_{G,G,G,G}&=-6 \hat{\lambda }_3-9 \hat{\Lambda }_6 \varphi ^2, \\
C_{G,G,G^+,G^-}&=-2 \hat{\lambda }_3-3 \hat{\Lambda }_6 \varphi ^2, \\
C_{G^+,G^+,G^-,G^-}&=-4 \hat{\lambda }_3-6 \hat{\Lambda }_6 \varphi ^2.
\end{align}

Using these, we may calculate the effective potential to two-loop level in the SM-like + 6-dim. effective theory: the tree-level contribution reads 
\begin{align}
V^{(0)}_\text{eff} = \frac12 \hat{\mu}^2_3 \varphi^2 + \frac14 \hat{\lambda}_3 \varphi^4 + \frac18 \hat{\Lambda}_6 \varphi^6,
\end{align}  
while the required integrals for one- and two-loop corrections have been presented in Appendix B.2 of Ref.~\cite{Farakos:1994kx}. At two-loop order, the 6-dim. operator enters the calculation only via its contribution to the masses and couplings listed above, so generalizing the calculation of Ref.~\cite{Farakos:1994kx} for our purposes is straightforward and will not be presented explicitly. For the analysis in the companion paper \cite{Andersen:2017ika}, we have discarded the $\gr{U(1)}$ coupling $\hat{g}'_3$, which has little effect on our 6-dim. error estimate.   

Due to the presence of the 6-dim. operator, the effective theory is no longer super-renormalizable. UV divergences arising at two-loop level can be canceled by introducing the following counterterms in the tree-level part:
\begin{align}
&\delta\hat{\mu}_3^2 = -\frac{1}{16\pi^2}\frac{1}{4\epsilon}\Big( \frac{51}{16}\hat{g}_3^4 + 9\hat{g}_3^2 \hat{\lambda}_3 - 12 \hat{\lambda}_3^2 \Big), \\
&\delta\hat{\lambda}_3 = -\frac{1}{16\pi^2}\frac{1}{2\epsilon} (9\hat{\Lambda}_6 \hat{g}_3^2 - 48 \hat{\Lambda}_6 \hat{\lambda}_3), \\ 
&\delta \hat{\Lambda}_6 = \frac{1}{16\pi^2}\frac{1}{\epsilon} 51 \hat{\Lambda}^2_6.
\end{align}
 
\newpage

\bibliographystyle{JHEP}

\end{document}